# Tip-Based Proximity Ferroelectric Switching and Piezoelectric Response in Wurtzite Multilayers

Eugene A. Eliseev[1*], Anna N. Morozovska[2†], Sergei V. Kalinin[3‡], Long-Qing Chen[4§], and Venkatraman Gopalan[4**]

[1]Frantsevich Institute for Problems in Materials Science, National Academy of Sciences of Ukraine, 3, str. Omeliana Pritsaka, 03142 Kyiv, Ukraine

[2] Institute of Physics, National Academy of Sciences of Ukraine, 46, Nauki avenue, 03028 Kyiv, Ukraine

[3]Department of Materials Science and Engineering, University of Tennessee, Knoxville, TN, 37996, USA

[4] Department of Materials Science and Engineering, Pennsylvania State University, University Park, PA 16802, USA

## Abstract

Proximity ferroelectricity is a novel paradigm for inducing ferroelectricity, when a non-ferroelectric polar material (such as AlN), which is unswitchable with an external field below the dielectric breakdown field, becomes a practically switchable ferroelectric in direct contact with a thin switchable ferroelectric layer (such as $Al_{1-x}Sc_xN$). Here, we develop a Landau-Ginzburg-Devonshire approach to study the proximity effect of local piezoelectric response and polarization reversal in wurtzite ferroelectric multilayers under a sharp electrically biased tip. Using finite element modeling we analyze the probe-induced nucleation of nanodomains, the features of local polarization hysteresis loops and coercive fields in the $Al_{1-x}Sc_xN$/AlN bilayers and three-layers. Similar to the wurtzite multilayers sandwiched between two parallel electrodes, the regimes of "proximity switching" (when all layers collectively switch) and the regime of "proximity suppression" (when they collectively do not switch) are the only two possible regimes in the probe-electrode geometry. However, the parameters and asymmetry of the local piezo-response and polarization hysteresis loops depend significantly on the sequence of the layers with respect to the probe. The physical mechanism of the proximity ferroelectricity in the local probe geometry is a depolarizing electric field determined by the polarization of the layers and their relative thickness. The field, whose direction is opposite to the polarization vector in the layer(s) with the larger spontaneous polarization (such as AlN), renormalizes the double-well ferroelectric potential to lower the steepness of the switching barrier in the "otherwise unswitchable" polar layers. Tip-based control of domains in otherwise non-ferroelectric layers using proximity ferroelectricity can provide nanoscale control of domain reversal in memory, actuation, sensing and optical applications. The ability of the tip-induced proximity switching to differentially switch multilayers, based on the order of the layers, provides a powerful knob for selective domain engineering.

---

[*]These authors contributed equally

[†]These authors contributed equally

[‡] corresponding author, e-mail: sergei2@utk.edu

[§] corresponding author, e-mail: lqc3@psu.edu

[**] corresponding author, e-mail: vgopalan@psu.edu

## 1. Introduction

"Proximity ferroelectricity" was recently reported by Skidmore et al. [1] who define this phenomenon as follows: "Proximity ferroelectricity is an interface-associated phenomenon, where electric field driven polarization reversal in a non-ferroelectric polar material is induced by one or more adjacent ferroelectric materials". The proximity ferroelectricity was revealed experimentally in the non-ferroelectric layers (such as AlN and ZnO) coupled with the ferroelectric layers (such as $Al_{1-x}B_xN$, $Al_{1-x}Sc_xN$ and $Zn_{1-x}Mg_xO$) in nitride-nitride, oxide-oxide, and nitride-oxide multilayers. The layered structures, whose thicknesses varied from tens to hundreds of nm, included two-layer (asymmetric, e.g. $Al_{1-x}Sc_xN$/AlN, $Al_{1-x}B_xN$/AlN, ZnO/$Al_{1-x}B_xN$) and three-layer (symmetric, e.g. $Al_{1-x}B_xN$/AlN/$Al_{1-x}B_xN$, AlN/$Al_{1-x}B_xN$/AlN, $Zn_{1-x}Mg_xO$/ZnO/$Zn_{1-x}Mg_xO$) configurations [1]. The tip-based proximity ferroelectric reversal will potentially provide local control of domain structures, which is important for memory, piezoelectric and optical applications.

We note that the ferroelectric layers such as $Zn_{1-x}Mg_xO$, $Al_{x-1}Sc_xN$ and $Al_{x-1}B_xN$ in the examples above, as well as others such as $Hf_xZr_{1-x}O_2$, rely on doping-induced chemical stress to turn a non-ferroelectric layer into a ferroelectric layer: Zr doping of $HfO_2$, Mg doping of polar but unswitchable ZnO, and Sc or B doping of the polar but unswitchable AlN) [2, 3, 4, 5, 6]. The local lattice distortions and charged defects around chemical dopants can locally allow for polarization switching to create domain nuclei, which then propagate through long-range electrostatic forces [7, 8]. These materials have become promising candidates for the next-generation of Si-compatible electronic memory elements such as ferroelectric random-access memories (FeRAMs), steep-slope field-effect transistors (FETs), and other logic devices [9, 10, 11, 12, 13]. However, the parent phases (AlN and ZnO) are not switchable with practical electric fields unless the proximity effect is exploited in a multilayer film stack.

The goal of this work is to develop the theoretical formalism for proximity switching and piezoelectric response in ferroelectric/non-ferroelectric multilayers under a sharp, electrically biased tip such as those used in scanning probe microscopes. Theoretical calculations of the polar properties and polarization switching in alternating ferroelectric and paraelectric (or dielectric) bilayers, multilayers and superlattices, placed in the capacitor geometry, were performed using the Landau-Ginzburg-Devonshire (LGD) phenomenological approach [14, 15, 16, 17], the first principles [18, 19], combination of the LGD approach with the phase-field method [20] and/or with direct variational methods [21, 22, 23]. These works, as very many others, consider perovskite-perovskite bilayers, multilayers and superlattices, such as $BaTiO_3/SrTiO_3$, $PbTiO_3/SrTiO_3$, $CaTiO_3/BaTiO_3$ and $BiFeO_3/SmFeO_3$.

Eliseev et. al. [24] have developed a Landau-Ginzburg theory of proximity ferroelectricity in wurtzite multilayers of non-ferroelectrics and ferroelectrics to analyze their switchability and coercive fields. The theory predicts regimes of both "proximity switching", where the multilayers collectively switch, as well as "proximity suppression" where they collectively do not switch. The mechanism of the proximity ferroelectricity is an internal electric field determined by the polarization of the layers and their relative thickness in a self-consistent manner that renormalizes the double-well ferroelectric potential to lower the steepness of the switching barrier. Further reduction in the coercive field emerges from charged defects in the bulk that act as nucleation centers, since correlated nucleation of the spike-like domains in the vicinity of sign-alternating randomly distributed electric charge sources [25], as well as correlated polarization switching in the proximity of ferroelectric domain walls [26], have a significant influence on the coercive field reduction.

This theoretical work tests the hypothesis that a biased piezoelectric force microscope (PFM) probe can be considered as the strong external charged defect, and so the strong reduction of the coercive field can be expected in the case of local polarization reversal under the probe. The local piezoelectric response (PFM signal) can be simulated in the semi-analytical linear decoupling approximation [27, 28, 29], partial coupling between the ferroelectric polarization and elastic fields [30], or by solving numerically the fully coupled problem [31, 32]. The simulation results obtained in the decoupling approximation well describe the results of PFM experiments (see e.g., a topical review [33] and refs. therein).

There are several methods for simulations of probe-induced nucleation of nanodomains, which include the semi-analytical approach proposed by Prof. M. Molotskii group [34, 35, 36], the phase-field simulations performed by Prof. L.-Q. Chen group [37, 38, 39], hibrid of mesoscopic and microscopic approach created by Prof. A. Rappe group [40, 41, 42]. The full list, of couse, is far not limited by these works.

In this work we developed the LGD approach to study the proximity effect of local polarization reversal and piezoelectric response in wurtzite ferroelectric multilayers. Using the finite element modeling (FEM) we analyze the probe-induced nucleation of nanodomains, the features of local polarization hysteresis loops and coercive fields in the $Al_{1-x}Sc_xN$/AlN bilayers and three-layers. We reveal that the parameters and asymmetry of the local piezoelectric response and polarization hysteresis loops depend significantly on the sequence of the layers with respect to the probe (for example whether the ferroelectric $Al_{1-x}Sc_xN$ or the non-ferroelectric AlN layer is directly in contact with the probe).

## 2. Problem Statement

**Figures 1(a)-(d)** depict the considered geometries of asymmetric bilayers and symmetric three-layers, consisting of AlN and $Al_{1-x}Sc_xN$ layers placed between the PFM probe apex and the bottom flat electrode. The total thickness of $Al_{1-x}Sc_xN$ layers is $h_1$ and the total thickness of the AlN layers is $h_2$. The total thickness of the multilayer stack, $h$, is the same for all considered geometries.

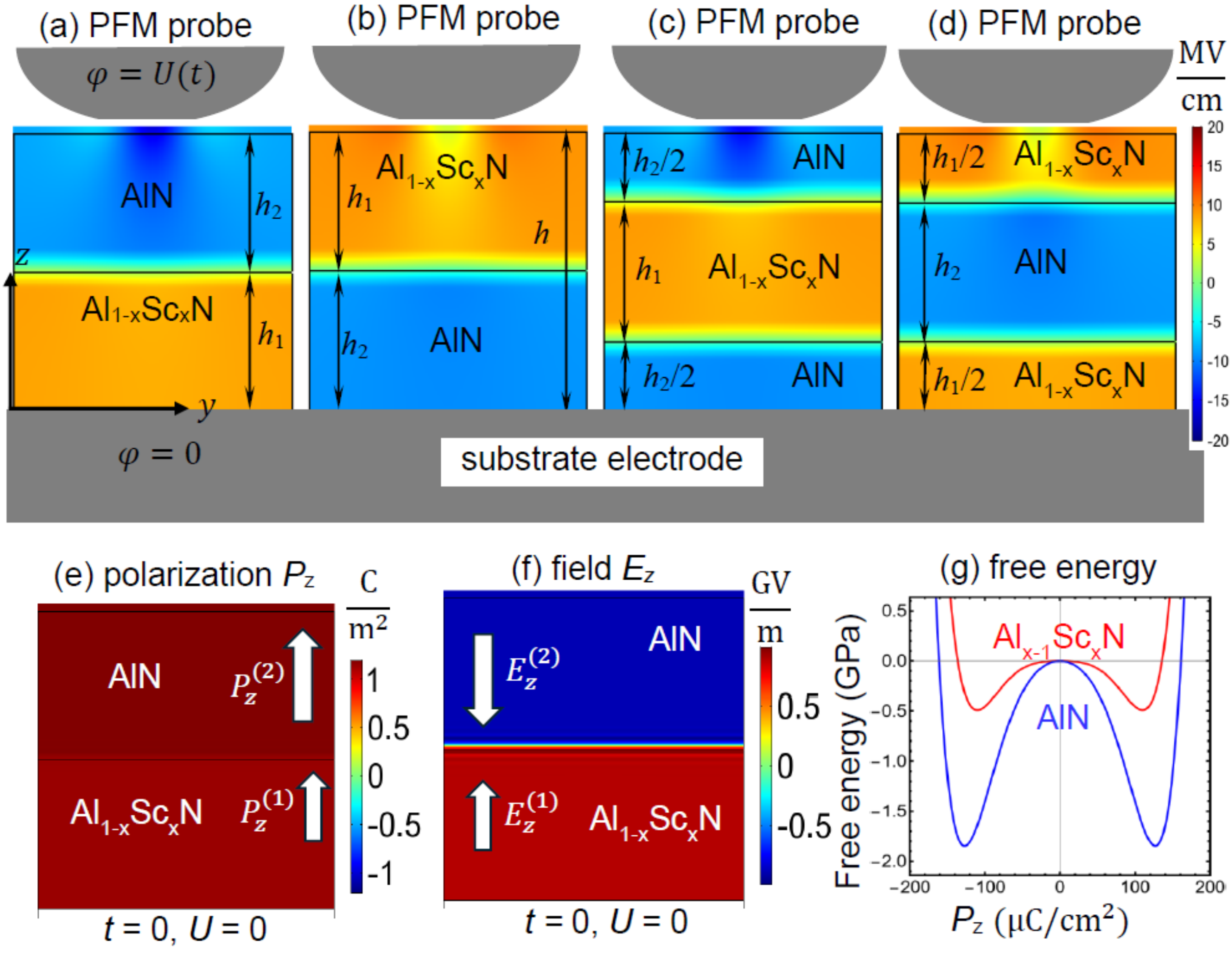

**FIGURE 1**. Geometries of considered heterostructures: the "probe/AlN/$Al_{1-x}Sc_xN$/electrode" **(a)**, the "probe/$Al_{1-x}Sc_xN$/AlN/electrode" **(b)**, the "probe/AlN/$Al_{1-x}Sc_xN$/AlN/electrode" **(c)**, and the "probe/$Al_{1-x}Sc_xN$/AlN/$Al_{1-x}Sc_xN$/electrode" **(d)**. The sample coordinates (x, y, z) form a right-handed coordinate system. The thickness of $Al_{1-x}Sc_xN$ layer is $h_1$ and the thickness of the AlN layer is $h_2$. Color maps show the electric field distributions for the considered geometries of single-domain multilayers for the bias $U \approx$ -7.5 V. The right scale corresponds to the electric field in MV/cm. The total thickness of the multilayer stack, $h = 40$ nm, is the same for all considered heterostructures. Initial distributions of the spontaneous polarization **(e)** and depolarization electric field **(f)** in the heterostructure "probe/AlN/$Al_{1-x}Sc_xN$/electrode". **(g)** Free energy wells of $Al_{0.73}Sc_{0.27}N$ (red curve) and AlN (blue curve) bulk materials.

The tip-surface contact is considered using the model of "shielded probe" [43, 44] that allows one to fix the electric potential $\varphi$ at the unperturbed surface $z = h$. Within the model, the PFM probe apex is approximated by a biased disk of the radius $R$. The free surface of the wurtzite layer outside the disk electrode is covered by a layer of screening charge, which is capable to provide the condition $\varphi = 0$ outside the probe-surface contact. The model is mostly realistic and accurate, because "squashed" tips are widely used, being such as prepared or become squashed during the exploitation. The accuracy of the approximation for a truncated-conical shape probe, which electric potential can be rigorously represented by the charged disk (tip part) and the linear charge (conical part), is analyzed in Ref. [44]. The shielded probe model is well applicable for the squashed tips, if the surface screening of the ferroelectric polarization is present outside the tip-surface contact. Since the screening outside the tip-contact area is provided by the ambient charges, which may be fast or retarding depending on the ambient conditions (e.g., humidity), the relaxation time of polarization changes and on the sweep frequency, the natural limitation of the shielded probe model is the fast switching of polarization at high bias voltages, when the lateral size of the tip-induced domain well exceeds the radius $R$, and the ambient charges outside the tip-contact area do not have enough time to screen the bias-induced changes of the ferroelectric polarization. Below we do not consider such situations.

A time-dependent electric voltage $U(t)$ is applied to the shielded probe. The voltage pulse is sinusoidally modulated and its amplitude increasing linearly in time from zero to $U_{max}$, namely $U(t) = U_{max}\frac{t}{t_{max}}sin(\omega t)$, where $\omega$ is the pulse frequency and $t_{max}$ is the pulse duration. We performed the FEM for 2 nm$\leq R \leq$10 nm, that is significantly smaller than $h$~40 nm. The free surface of the wurtzite layer outside the disk electrode is covered by a layer of screening charges (not shown in **Fig. 1**), which is capable of providing the condition $\varphi = 0$ outside the probe-surface contact. The smooth transition between the edge of the biased disk-probe and the screened surface is modeled by a Gaussian function.

Importantly, the shielded probe is regarded as "elastically soft" in order not to perturb the local piezoelectric response of the wurtzite layers. The substrate electrode is regarded as electrically grounded ($\varphi = 0$ at $z = 0$) and "elastically rigid". The elastic and electric boundary conditions are periodic at the side boundaries of the computational cell.

The time-dependent LGD equations for the $Al_{1-x}Sc_xN$ layer (subscripts and superscript "1") and the AlN layer (subscripts and superscript "2") are the following [24]:

$$\Gamma_1\frac{\partial P_z^{(1)}}{\partial t} + \tilde{\alpha}_1 P_z^{(1)} + \beta_1\left(P_z^{(1)}\right)^3 + \gamma_1\left(P_z^{(1)}\right)^5 - g_z^{(1)}\frac{\partial^2 P_z^{(1)}}{\partial z^2} - g_\perp^{(1)}\left(\frac{\partial^2 P_z^{(1)}}{\partial x^2} + \frac{\partial^2 P_z^{(1)}}{\partial y^2}\right) = E_z^{(1)}, \quad (1a)$$

$$\Gamma_2\frac{\partial P_z^{(2)}}{\partial t} + \tilde{\alpha}_2 P_z^{(2)} + \beta_2\left(P_z^{(2)}\right)^3 + \gamma_2\left(P_z^{(2)}\right)^5 - g_z^{(2)}\frac{\partial^2 P_z^{(2)}}{\partial z^2} - g_\perp^{(2)}\left(\frac{\partial^2 P_z^{(2)}}{\partial x^2} + \frac{\partial^2 P_z^{(2)}}{\partial y^2}\right) = E_z^{(2)}. \quad (1b)$$

Here $\Gamma_i$ are the Landau-Khalatnikov relaxation coefficient of the layer "$i$", and we regard that $\Gamma_1 = \Gamma_2$ for the sake of simplicity. The coefficients $\tilde{\alpha}_i$, $\beta_i$ and $\gamma_i$ are the Landau expansion coefficients for $Al_{1-}$ ${}_{x}Sc_{x}N$ and AlN layers, and $E_z^{(i)}$ are the $z$-components of the electric field acting inside the layers. Hereafter, $i = 1$ corresponds to $Al_{1-x}Sc_xN$ layer(s) and $i = 2$ corresponds to AlN layer(s). LGD parameters of $Al_{1-x}Sc_xN$ and AlN, used in our calculations, are listed in **Table I.** They were determined from the experimentally measured spontaneous polarization [45, 46] and linear dielectric permittivity [47] as described in Refs. [48] and [24]. Free energy wells of $Al_{0.73}Sc_{0.27}N$ and AlN bulk materials are shown in **Fig. 1(g).**

The boundary conditions for the polarization at the top surface, interfaces and bottom surface of the multilayer stack are of the third type [49, 50] in a general case (i.e., a function plus its derivative is a constant; see Eqs.(3) in Ref. [24] for details). Below, for the sake of simplicity, we regard that the polarization components and their derivatives are continuous at the interfaces between AlN and $Al_{1-x}Sc_xN$ layers. The boundary conditions at the top and bottom surfaces are regarded "natural", namely $\frac{\partial P_z^{(i)}}{\partial z} = 0\Big|_{z=0}$ and $\frac{\partial P_z^{(i)}}{\partial z} = 0\Big|_{z=h}$. The initial distribution of polarization is either a single-domain or a multi-domain state with randomly small fluctuations.

The coefficients $\tilde{\alpha}_i$ in Eqs.(1) are renormalized by elastic stresses $\sigma_{kl}^{(i)}$ as $\tilde{\alpha}_i = \alpha_i - Q_{12}^{(i)}\left(\sigma_{22}^{(i)} + \sigma_{33}^{(i)}\right) - Q_{11}^{(i)}\sigma_{11}^{(i)}$, where $Q_{kl}^{(i)}$ are electrostriction coefficients of the layers, which are listed in **Table II**. Elastic stresses satisfy the equation of mechanical equilibrium in the layers.

$$\frac{\partial \sigma_{jk}^{(i)}}{\partial x_j} = 0. \tag{2a}$$

Elastic equations of state follow from the variation of the free energy with respect to elastic stress, namely:

$$s_{jklm}^{(i)}\sigma_{lm}^{(i)} + Q_{jklm}^{(i)}P_l^{(i)}P_m^{(i)} = u_{jk}^{(i)}. \tag{2b}$$

Here $u_{jk}^{(i)}$ are elastic strains, and $s_{jklm}^{(i)}$ are elastic compliances (also listed in **Table II**).

Elastic boundary conditions correspond to the absence of normal stresses at the top surface $z = h$, elastic displacement continuity at the AlN – $Al_{1-x}Sc_xN$ interface (or interfaces), and zero elastic displacement in the rigid substrate, $z = 0$. Hereafter, we neglect the influence of the flexoelectric coupling for the sake of simplicity.

Electric field in the layers, $E_z^{(i)} = -\frac{\partial \varphi^{(i)}}{\partial z}$, obeys the Poisson equations for the electric potential $\varphi^{(i)}$ inside the layers:

$$\varepsilon_0\varepsilon_b^{(i)}\left(\frac{\partial^2}{\partial x^2} + \frac{\partial^2}{\partial y^2} + \frac{\partial^2}{\partial z^2}\right)\varphi^{(i)} = \frac{\partial^2 P_z^{(i)}}{\partial z^2}. \tag{3}$$

Here $\varepsilon_b^{(i)}$ is the background dielectric permittivity [51, 52] of the layer "$i$". The electric boundary conditions are the continuity of $\varphi^{(i)}$ and electric displacement component $D_z^{(i)}$ at the AlN – $Al_{1-x}Sc_xN$ interfaces, and the fixed potential at the electrodes, e.g., $\varphi^{(1)} = 0\big|_{z=0}$, $\varphi^{(2)} = U(t)exp\left(-\frac{x^2+y^2}{R^2}\right)\Big|_{z=h}$.

**Table I.** LGD model parameters*

| compound | $\alpha_i$, m/F | $\beta_i$, m$^5$/(F C$^2$) | $\gamma_i$, m$^7$/(F C$^4$) | $g_z^{(i)}$, m$^3$/F | $g_\perp^{(i)}$, m$^3$/F | $\varepsilon_b^{(i)}$ |
|---|---|---|---|---|---|---|
| $Al_{0.73}Sc_{0.27}N$ | $-2.792 \cdot 10^{8}$** | $-3.155 \cdot 10^{9}$ | $2.788 \cdot 10^{9}$ | $5 \cdot 10^{-10}$ | $1 \cdot 10^{-10}$ | 3** |
| AlN | $-2.164 \cdot 10^{9}$ | $-3.155 \cdot 10^{9}$ | $2.788 \cdot 10^{9}$ | $5 \cdot 10^{-10}$ | $1 \cdot 10^{-10}$ | 4 |

* The parameters are taken from Table CII, given in the supplemental to Ref. [24], except those marked with "**"

**The parameter $\alpha_1$ is refined in this work allowing for the electrostriction energy contribution; the background permittivity $\varepsilon_b^{(1)}$ is not equal to $\varepsilon_b^{(2)}$, because the approximation $\varepsilon_b^{(1)} \approx \varepsilon_b^{(2)}$, used in Ref. [24], is not used hereinafter.

**Table II.** Electrostriction coefficients $Q_{ij}$ and elastic stiffness $c_{ij}$

| parameters | $Q_{ij}$, m$^4$/C$^2$ | Ref. | $c_{ij}$, GPa | Ref. |
|---|---|---|---|---|
| AlN | $Q_{13} = -0.0087$, $Q_{33} = 0.0203$ | [48] | $c_{11}$=396, $c_{12}$=137, $c_{13} = 108$, $c_{33} = 373$, $c_{44} = 116$, $c_{66} = 130$ | [53] |
| $Al_{0.73}Sc_{0.27}N$ | $Q_{13} = -0.0152$, $Q_{33} = 0.0406$ | [48] | $c_{11}$=319, $c_{12}$=151, $c_{13} = 127$, $c_{33} = 249$, $c_{44} = 101$, $c_{66} = 84$ | [53] |

Despite the fact that the bias applied between the tip and the flat bottom electrode is the same ($U = -7.5$ V) in all four cases shown in **Fig. 1(a)-(d)**, and the spontaneous polarization of the single-domain layers is directed upward (see e.g., **Fig. 1(e)**), the sign of electric field is opposite in the AlN and $Al_{1-x}Sc_xN$ layers due to the dominant depolarization field effect, which is "inverted" in the AlN layer (see e.g., **Fig. 1(f)**).

The effect of the field "inversion" can be described analytically when domains are absent, and the polarization is almost constant inside each layer. Indeed, when $\frac{\partial P_z^{(i)}}{\partial z} = 0$, the depolarization field inside the multilayer, whose top and bottom surfaces are covered by conducting electrodes, is given by the analytical expression derived in Ref.[24]. These expressions allow us to estimate the electric field in the shielded probe geometry if $2R \sim h$. The approximate expressions for the electric field inside the bilayer are:

$$E_z^{(1)}(x,y,z,t) \approx -\frac{P_z^{(1)}-\bar{D}}{\varepsilon_0 \varepsilon_b^{(1)}} + \frac{U(t)exp\left(-\frac{x^2+y^2}{R^2}\right)R^2}{\varepsilon_b^{(1)}\left[\left(h_1/\varepsilon_b^{(1)}\right)+\left(h_2/\varepsilon_b^{(2)}\right)\right]\left(R+z/\chi_b^{(1)}\right)^2}, \tag{3a}$$

$$E_z^{(2)}(x,y,z,t) \approx -\frac{P_z^{(2)}-\overline{D}}{\varepsilon_0\varepsilon_b^{(2)}} + \frac{U(t)exp\left(-\frac{x^2+y^2}{R^2}\right)R^2}{\varepsilon_b^{(2)}\left[\left(h_1/\varepsilon_b^{(1)}\right)+\left(h_2/\varepsilon_b^{(2)}\right)\right]\left(R+z/\chi_b^{(2)}\right)^2}. \quad (3b)$$

The first term in Eq. (3), $-\frac{P_z^{(i)}-\overline{D}}{\varepsilon_0\varepsilon_b^{(i)}}$, is the internal depolarization field $E_{dz}^{(i)}$ [54, 55]. The second term in Eq. (3) is the external field proportional to the probe bias $U(t)$ [28]. The values $\chi_b^{(i)}$ are dielectric anisotropy factor of the layers. $\overline{D}$ is the average displacement of the multilayer, which is equal to [56]:

$$\overline{D} = \frac{1}{\left(h_1/\varepsilon_b^{(1)}\right)+\left(h_2/\varepsilon_b^{(2)}\right)}\left(\frac{h_1}{\varepsilon_b^{(1)}}\overline{P}_z^{(1)} + \frac{h_2}{\varepsilon_b^{(2)}}\overline{P}_z^{(2)}\right), \quad (4)$$

where $\overline{P}_z^{(i)}$ is the average polarization of the layer "$i$". Note that the average displacement $\overline{D}$ coincides with the average polarization $\overline{P}$ when $\varepsilon_b^{(i)}$ is the same for both layers. From Eqs. (3)-(4), the depolarization fields are opposite, namely $E_{dz}^{(1)} = -E_{dz}^{(2)} = -\frac{P_z^{(1)}-P_z^{(2)}}{2\varepsilon_0\varepsilon_b}$, for $h_1 = h_2 = \frac{h}{2}$ and $\varepsilon_b^{(1)} = \varepsilon_b^{(2)} = \varepsilon_b$. Since the depolarization field is much higher than the probe field at small biases, the simple estimate explains the field inversion effect shown in **Fig. 1(f)**. From **Fig. 1(f)**, the depolarization field of the AlN/$Al_{1-x}Sc_xN$ bilayer with polarization directed upward in both layers (as shown in **Fig. 1(e)**) is directed downward in the AlN layer and upward in the $Al_{1-x}Sc_xN$ layer.

Below we will show that the opposite direction of the field in the AlN and $Al_{1-x}Sc_xN$ layers, as well as its strong inhomogeneity under the biased probe, are the main reasons for the dependence of local polarization reversal and piezoelectric response on the sequence of the layers. The depolarization field, whose direction is opposite in the layer(s) with the larger spontaneous polarization (i.e., in the AlN), renormalizes the double-well ferroelectric potential to lower the steepness of the switching barrier in the otherwise "unswitchable" polar layers. Note that this effect is intrinsic to the multilayer structure and is independent of any additional reduction in coercive field due to the presence of defects.

The vertical surface displacement $u_z(x,y,t)$ is measured by the PFM, being directly related with the local piezoelectric response signal $d_{33}^{eff}$ as $d_{33}^{eff} = \frac{du_z}{dU} \sim \frac{u_z}{U}$ [29]. The linear relation $d_{33}^{eff} \sim \frac{u_z}{U}$ allows us to analyze the peculiarizes of $d_{33}^{eff}$ using the FEM results for $u_z$.

## 3. Results and Discussion

### A. Peculiarities of local polarization switching and piezoelectric response in bilayers and three-layers

Due to the inhomogeneous electric field distribution in the considered probe-electrode geometries (shown in **Figs. 1(a)-(d)**), the average polarization and local piezoelectric response depend on the sequence of the layers, being different for AlN/$Al_{1-x}Sc_xN$ (**Fig. 2**) and $Al_{1-x}Sc_xN$/AlN (**Fig. 3**) bilayers, as well as for AlN/$Al_{1-x}Sc_xN$/AlN (**Fig. 4**) and $Al_{1-x}Sc_xN$/AlN/$Al_{1-x}Sc_xN$ (**Fig. 5**) three-layers, where either AlN or $Al_{1-x}Sc_xN$ layer is in contact with the probe tip. Results, shown in various panels of **Figs. 2-5**, are calculated in the case of a single-domain initial distribution of polarization in all layers. Specifically, the initial state was an upward-directed spontaneous polarization with random small fluctuations in all layers. We waited until the initial relaxed to a single-domain upward-directed spontaneous polarization without any fluctuations (such as that shown in **Fig. 1(e)**) and then applied the bias according to the same timing protocol shown in **Figs. 2(a), 3(a), 4(a)** and **5(a)**. Note that the single-domain state of the spontaneous polarization is the ground state of the layered structure, since the surface screening is regarded ideal, and the boundary condition $\varphi = 0$ is valid at both surfaces.

The bias amplitude linearly increases in time as shown in **Figs. 2(a)-5(a)**. The time sweeps of the vertical surface displacement $u_z(0,0,t)$ under the PFM tip, which correspond to the bias sweep, are shown in **Fig. 2(b)-5(b)**. The time sweeps of the polarization *z*-component $\bar{P}_z^{aver}$ averaged over the volume of the computation cell, namely in the cylinder $\{0 \leq \sqrt{x^2 + y^2} < 4R,\ 0 \leq z \leq h\}$ with the volume $V_c \cong 16\pi R^2 h$, are shown in **Fig. 2(c)-5(c).** The butterfly-like loops of $u_z(0,0,t)$ are shown in **Figs. 2(d)-5(d)**. The ferroelectric hysteresis loops of the "average" polarization $\bar{P}_z^{aver}$, averaged over the cell volume $V_c$, are shown in **Figs. 2(e)-5(e)**. The ferroelectric hysteresis loops of the "local" polarization $\bar{P}_z^{local}$ averaged over a small region under the tip, namely in the cylinder $\{0 \leq \sqrt{x^2 + y^2} < R,\ 0 \leq z \leq h\}$ with the volume $V_t \cong \pi R^2 h$, are shown in **Figs. 2(f)-5(f)**. The distribution of polarization *z*-component $P_z$ in the cross-section of the bilayers and three-layers at the moments of time numbered from "1" to "10" in **Fig. 2(a)-5(a)** are shown in **Figs. 2(g)-5(g)**. The distribution of electric field at the moments "1" – "10" in the bilayers are shown in **Figs. S1** and **S2** in **Supplementary Materials** [57].

The local polarization reversal in bilayer and three-layer structures has several features, such as different positive and negative coercive bias, step-like features at the hysteresis loops and proximity switching effect in the AlN layer(s), which are listed in **Table III** and discussed in detail below.

**Table III.** Features of local polarization reversal in wurtzite multilayers

| Multilayer structure (figure number) | Positive and negative coercive biases ($U_c^-$ and $U_c^+$) *; the loop width | Step-like features (observed in the bias range 0 – 100 V) ** | Proximity switching effect in the AlN layer(s) (in dynamics) |
|---|---|---|---|
| $AlN/Al_{1-x}Sc_xN$<br>$h_{AlN} = h_{AlScN}$ =20 nm<br>(see **Fig. 2**) | $U_c^-$ ≈-23 V,<br>$U_c^+$ ≈21 V,<br>loop width ≈44 V | multiple step-like features are observed at negative bias | domain nucleation in the AlN layer and its vertical growth thorough the bilayer |
| $Al_{1-x}Sc_xN/AlN$<br>$h_{AlN} = h_{AlScN}$ =20 nm<br>(see **Fig. 3**) | $U_c^-$ ≈-45 V,<br>$U_c^+$ ≈32 V,<br>loop width ≈79 V | a single step-like features is observed | domain nucleation in the $Al_{1-x}Sc_xN$ layer and its vertical growth thorough the bilayer |

| AlN/$Al_{1-x}Sc_xN$/AlN<br>$2h_{AlN} = h_{AlScN}$ =20 nm<br>(see **Fig. 4**) | $U_c^-$ ≈-31 V,<br>$U_c^+$ ≈25 V,<br>loop width ≈56 V | multiple step-like features are observed at negative bias | domain nucleation in the AlN layer under the probe and its vertical growth thorough the layers |
|---|---|---|---|
| $Al_{1-x}Sc_xN$/AlN/$Al_{1-x}Sc_xN$<br>$h_{AlN} = 2h_{AlScN}$ =20 nm<br>(see **Fig. 5**) | $U_c^-$ ≈-38 V,<br>$U_c^+$ ≈32 V,<br>loop width ≈70 V | several step-like features are observed at negative bias | domain nucleation in the $Al_{1-x}Sc_xN$ layer under the probe and its vertical growth thorough the layers |

* $U_c^-$ and $U_c^+$ are the negative and positive coercive biases corresponding to the red loops in **Figs. 2-5**

**The direction of the spontaneous polarization in the layers determines the sign of the bias for which the pronounced step-like polarization response is observed. In the considered case the spontaneous polarization was directed upward in all layers when the bias was absent (see e.g., **Fig. 1(e)**).

The small bias amplitude (below the coercive voltage) is insufficient to induce the local polarization reversal under the probe (see black curves in **Fig. 2(e)-5(e)** and **Fig. 2(f)-5(f)**). Since the biased probe acts as a strong external charged defect, the switching of polarization is multi-domain even in the absence of point charge or elastic defects, which are not considered in this work. When the probe field overcomes the coercive field, the domain nucleus emerges under the probe and rapidly grows through the multilayer, leading to the switching within the smallest red loop nested inside the larger magenta and blue loops. The polarization loop opens with an increase in the bias amplitude (see magenta and blue loops in **Fig. 2(e)-5(e)** and **Fig. 2(f)-5(f)**). When the probe bias increases further, the domain walls start to move in the transverse direction; when the bias changes the sign, an oppositely polarized domain starts to grow inside the existing domain (see **Fig. 2(g)-5(g)**). We would like to emphasize that the up shift of $\bar{P}_z^{aver}$ loops, shown in **Figs. 2(e)-5(e)**, is conditioned by the polarization averaging over the large area $S_c \gg \pi R^2$; and the shift virtually disappears when the averaging is performed over the small tip-surface contact area $S_t \cong \pi R^2$ (see the loops of $\bar{P}_z^{local}$ in **Figs. 2(f)-5(f)**). Thus, the averaging of polarization over the whole computation cell $V_c$ underestimates significantly the complete switching of polarization, which occurs under the probe tip (as shown by blue domains in **Figs. 2(g)-5(g)**). After the averaging over $V_c \gg \pi R^2 h$, it looks like the switching of polarization is always partial (see **Figs. 2(e)-5(e)**), which is not the case under the probe (see **Figs. 2(f)-5(f)**). However, for the purposes of this work we also consider the polarization $\bar{P}_z^{aver}$ because it determines the charge density stored at the electrodes (since the latter is proportional to the volume-averaged electric displacement), as well as it correlates with the behavior of the local piezoelectric response recorded by PFM.

Indeed, the vertical surface displacement $u_z(0,0,t)$ under the PFM tip, calculated at the point $x = y = 0$ without any averaging and shown in **Figs. 2(d)-5(d)**, responses to the changes of polarization in the area larger than $0 \leq \sqrt{x^2 + y^2} < R$, because elastic fields are very long-range (as it was shown using the decoupling approximation for PFM response [58]). The loops of $u_z$ are strongly asymmetric with respect to the initial state $u_z = 0$ (vertical asymmetry) as well as to the bias axis $U = 0$ (horizontal asymmetry). The asymmetry and step-like features of $u_z$ (shown in **Figs. 2(d)-5(d)**) well correlate with the asymmetry and features of the polarization hysteresis loops $\bar{P}_z^{aver}$ and $\bar{P}_z^{local}$ (shown in **Fig. 2(e)-5(e)** and **Figs. 2(f)-5(f)**, respectively). At that the step-like features are the most pronounced at the hysteresis loops of $\bar{P}_z^{aver}$.

The step-like features at the hysteresis loops of $u_z$, $\bar{P}_z^{aver}$ and $\bar{P}_z^{local}$ are associated with the pairwise "annihilation" of domain walls being similar to Barkhausen jumps in ferromagnets [59] and Barkhausen pulses in ferroelectric-ferroelastics [60]. As can be seen from **Fig. 2(g)-5(g),** the step-like features appear at the loops when the small domain, which grows inside the larger domain, reaches the boundaries of the larger domain. The step-like features are present at the loops independent of the initial polarization state, but the amount and position of the features depend on the initial state.

Note that the direction of the spontaneous polarization in the layers determines the sign of the bias for which the pronounced step-like polarization response is observed. In the considered case the ground-state spontaneous polarization was directed upward in all layers when the bias was switched off (see **Fig. 1(e)**). Then the pronounced step-like polarization response is observed when a negative bias is applied. When the spontaneous polarization was directed downward in all layers, the pronounced step-like polarization response is observed after a positive bias is applied. When the initial distribution of spontaneous polarization is a multi-domain state with randomly small fluctuations, the steps can occur for positive and negative biases (compare **Figs. S3(c)** and **S4(c)** in **Supplementary Materials** [57]). However, the multi-domain state can be the ground state for incomplete screening conditions only.

Sometimes strongly up- or down-shifted loops of local piezoresponse are observed experimentally (see e.g., figure 3 in Ref. [5]). It is worth noting that the character (upward or downward shift) and magnitude (large or small shift) of the vertical asymmetry of the $\bar{P}_z^{aver}$ hysteresis loops depend also on the initial state (single-domain or multi-domain) of polarization in the layers and its direction (upward or downward) under the probe. For example, the $\bar{P}_z^{aver}$ loops are shifted upward when the initial state of polarization is single-domain, and the initial polarization is directed upward. The vertical shift becomes downward for the multi-domain initial state of the polarization (compare **Fig. 2(e)** and **3(e)** with **Fig. S3(c)** in **Supplementary Materials** [57]). It is seen from **Fig. S4(c)** that it is possible to minimize the vertical and horizontal asymmetry of $\bar{P}_z^{aver}$ hysteresis loops by the appropriate choice of the initial multi-domain polarization state. This result can be explained considering that new domains may not arise from the multi-domain initial state during the switching, since the switching occurs via the growth of existing domains. In addition, the internal depolarization field is suppressed due to the presence of domains [50, 55], and, accordingly, the electric interaction between the layers is weaker.

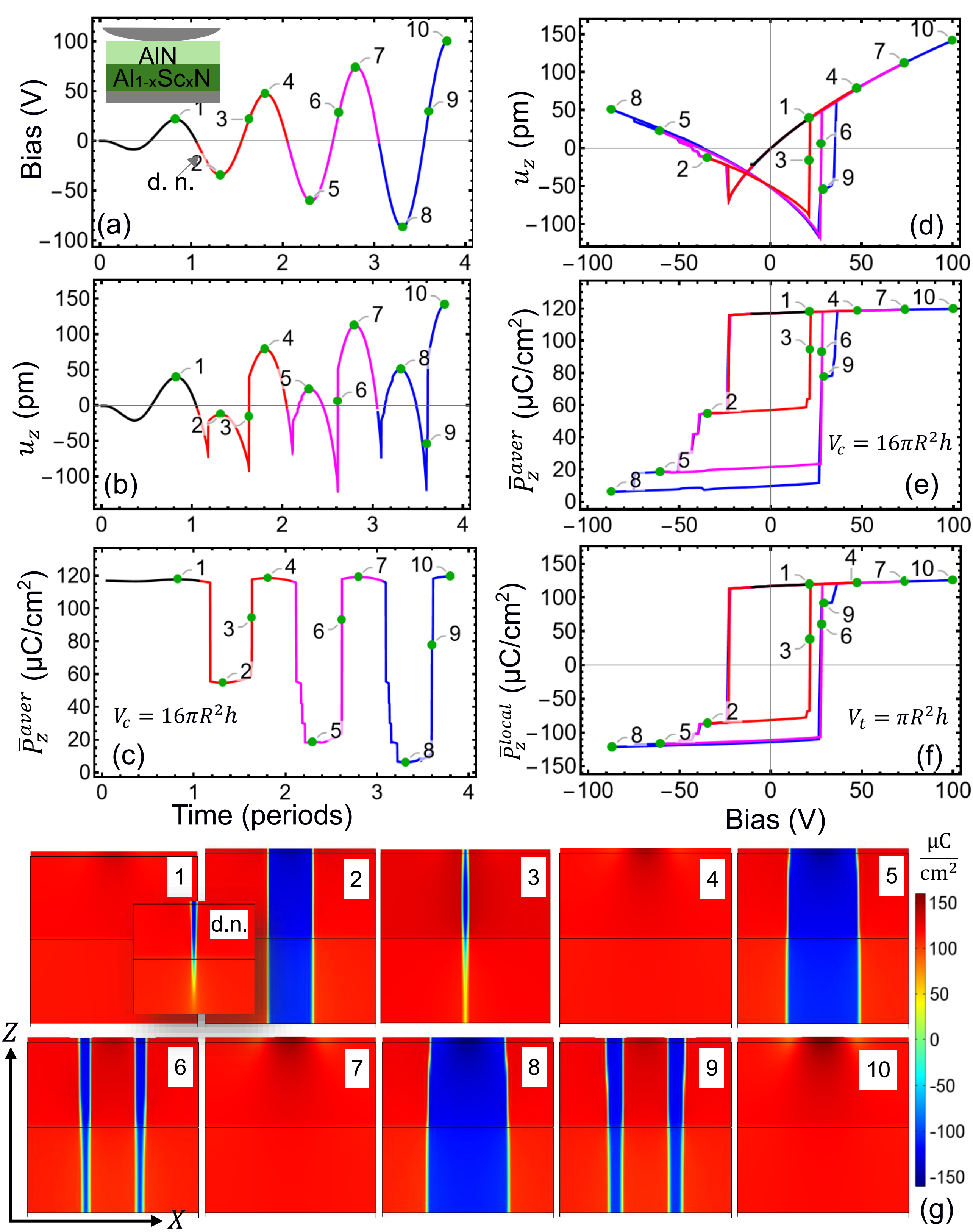

**FIGURE 2**. Time dependences of the bias applied between the PFM probe and the bottom electrode **(a)**, the vertical displacement $u_z$ of the surface below the probe tip **(b)** and the average polarization $\bar{P}_z^{aver}$ **(c)** in the heterostructure “probe

– AlN/$Al_{0.73}Sc_{0.27}N$ bilayer – bottom electrode". Bias dependences of the surface displacement $u_z$ **(d)**, the polarization $P_z$ averaged over a larger volume $V_c$ **(e)** and a smaller volume $V_t$ (**f**). The distribution of polarization **(g)** in the cross-section of the bilayer at the moments of time numbered from "1" to "10" shown by the pointers in (a) – (f). Abbreviation "d.n." means the domain nucleation. The distribution of electric field in the moments "1" – "10" is shown in **Figs. S1** (see **Supplementary Materials**). The thickness of the AlN and $Al_{0.73}Sc_{0.27}N$ layers is 20 nm, the tip-surface contact radius $R =$ 5 nm. LGD parameters and elastic constants are listed in **Tables I-II**. The averaging volumes $V_c = 16\pi R^2 h$ and $V_t = \pi R^2 h$; the initial distribution of polarization is a single-domain state with randomly small fluctuations.

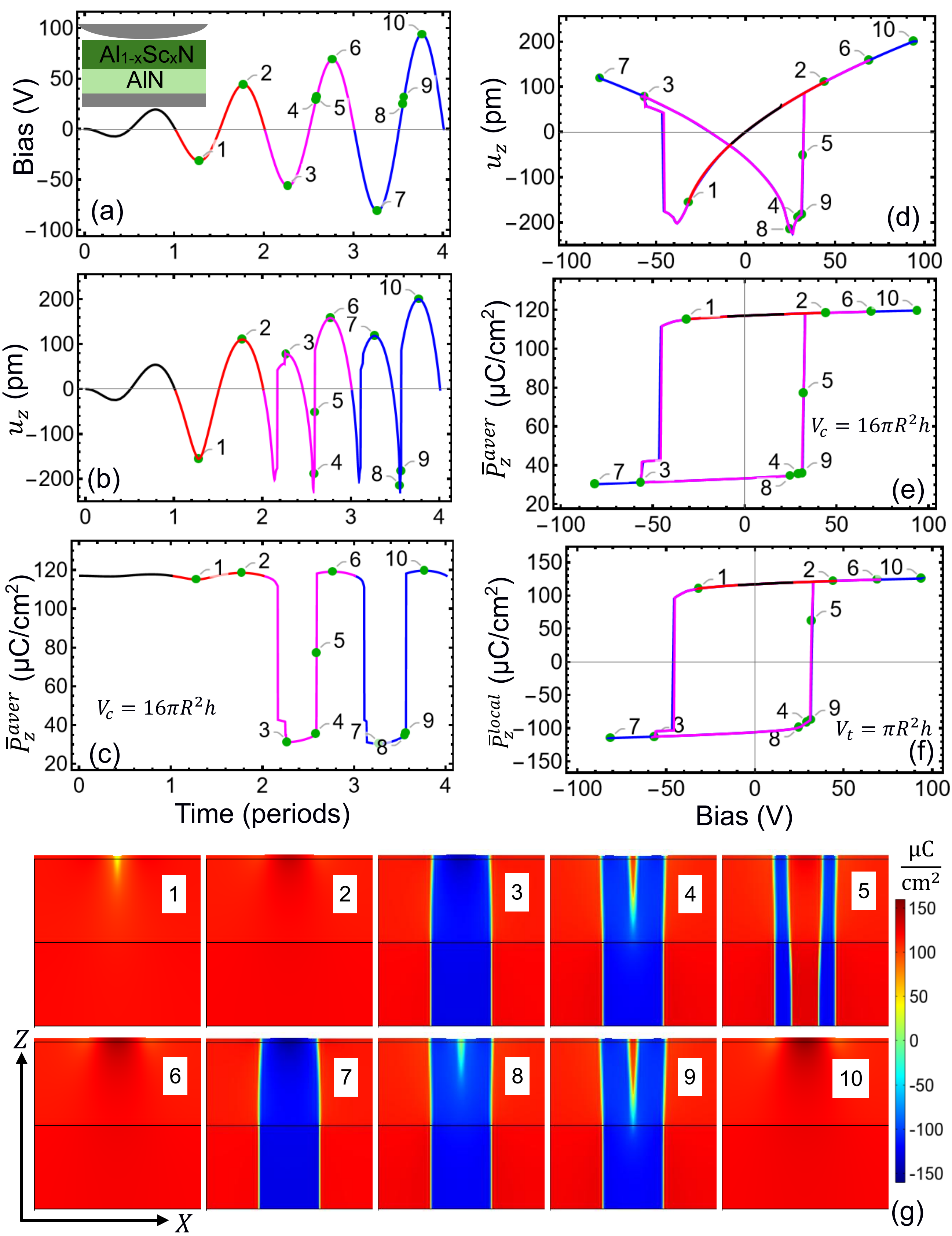

**FIGURE 3**. Time dependences of the bias applied between the PFM probe and the bottom electrode **(a)**, the vertical displacement $u_z$ of the surface below the probe apex **(b)** and the average polarization $\bar{P}_z^{aver}$ **(c)** in the heterostructure “probe – $Al_{0.73}Sc_{0.27}N$/AlN bilayer – bottom electrode”. Bias dependences of the surface displacement $u_z$ **(d)**, the polarization $P_z$ averaged over a larger volume $V_c$ **(e)** and a smaller volume $V_t$ **(f)**. The distribution of polarization **(g)** in the cross-section of the bilayer at the moments of time numbered from “1” to “10” shown by the pointers in (a) – (f). The distribution of electric field in the moments “1” – “10” is shown in **Figs. S2** (see **Supplementary Materials**). Other parameters and conditions are the same as in **Fig. 2.**

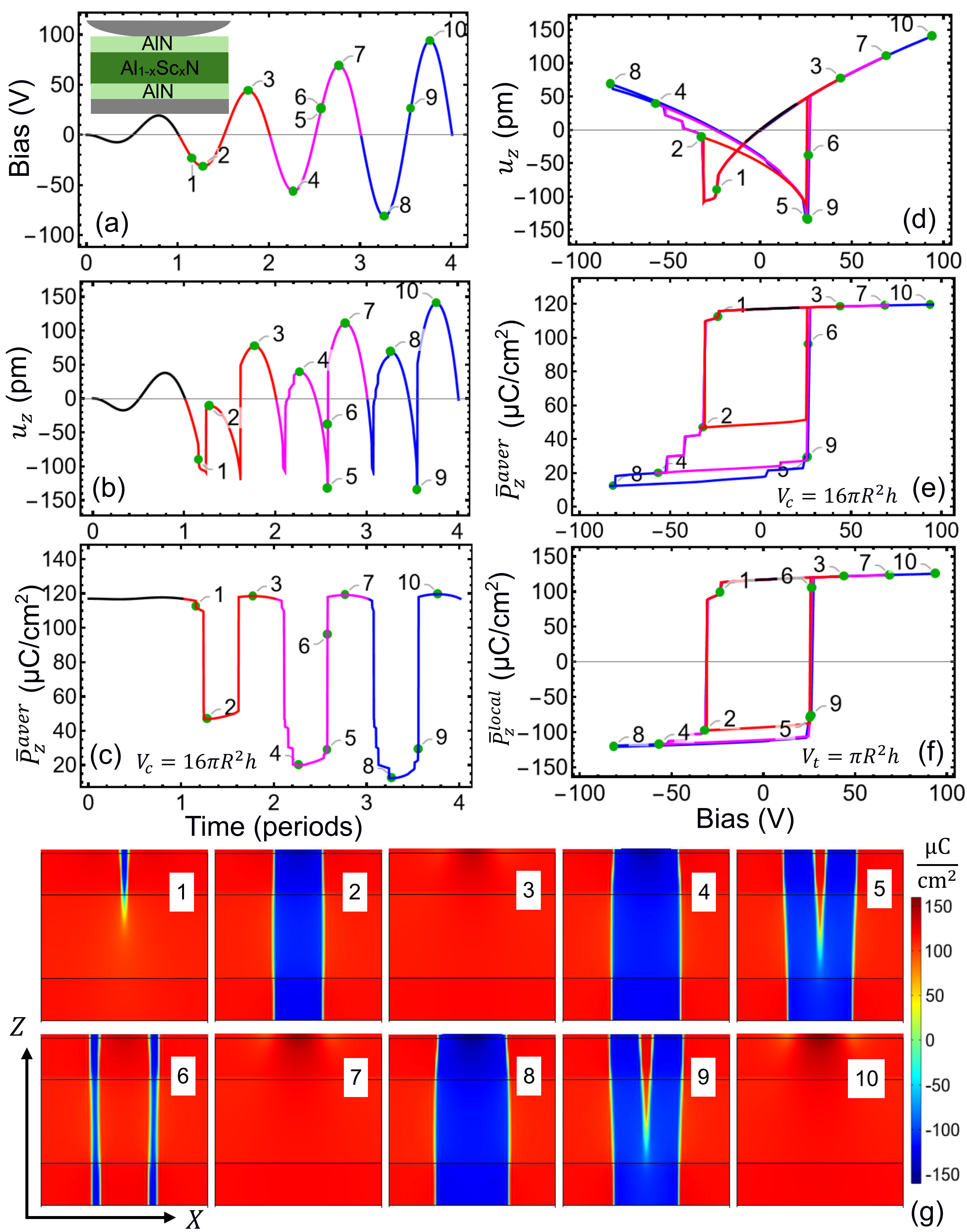

**FIGURE 4**. Time dependences of the bias applied between the PFM probe and the bottom electrode **(a)**, the vertical displacement $u_z$ of the surface below the probe apex **(b)** and the average polarization $\bar{P}_z^{aver}$ **(c)** in the heterostructure "probe – AlN/$Al_{0.73}Sc_{0.27}N$/AlN three-layer – bottom electrode". Bias dependences of the surface displacement $u_z$ **(d)**, the polarization $P_z$ averaged over a larger volume $V_c$ **(e)** and a smaller volume $V_t$ **(f)**. The distribution of polarization **(g)** in the cross-section of the three-layer at the moments of time numbered from "1" to "10" shown by the pointers in (a) – (f). The thickness of the $Al_{0.73}Sc_{0.27}N$ layer is 20 nm and the thickness of AlN layers is 10 nm, other parameters are conditions are the same as in **Fig. 2**.

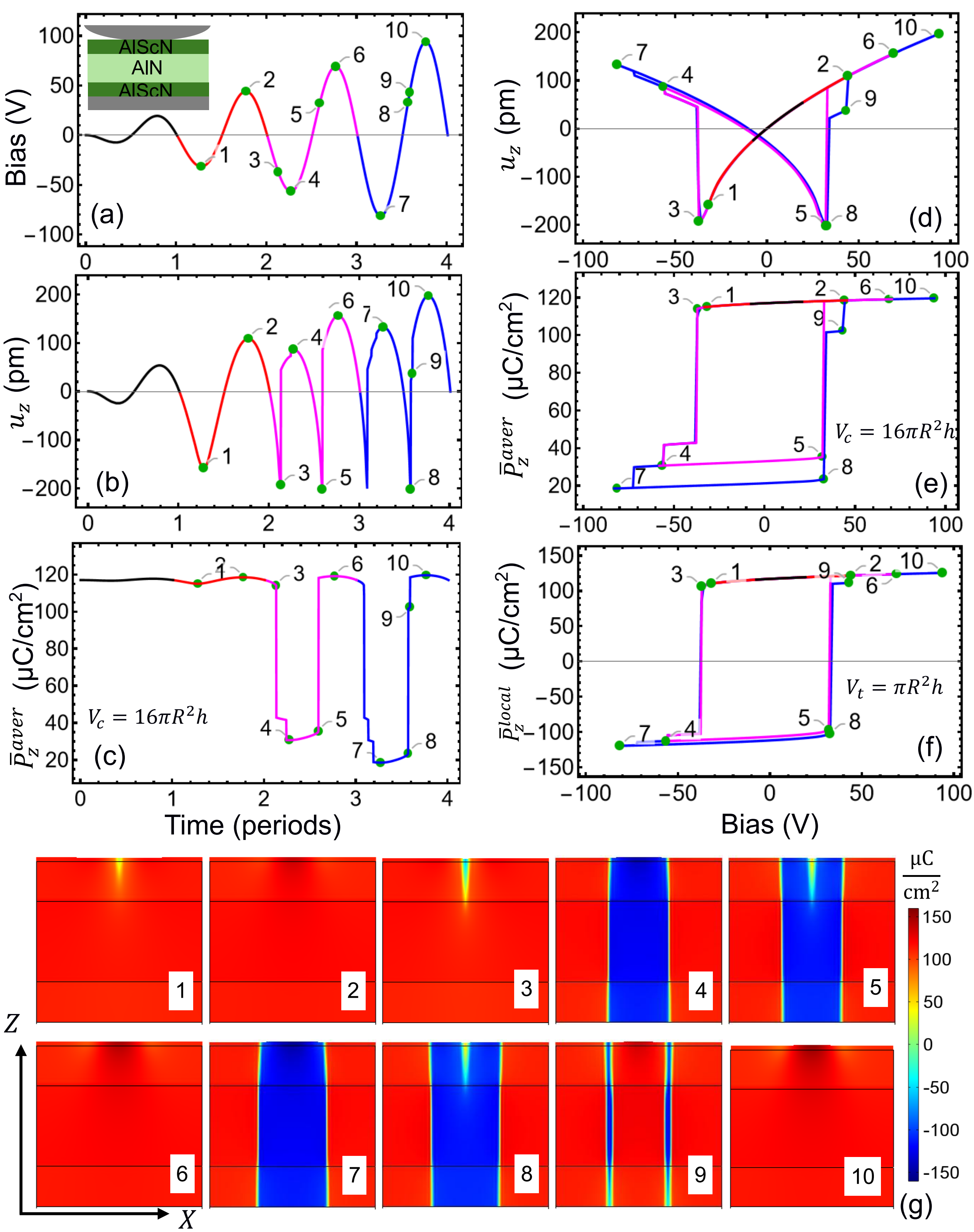

**FIGURE 5**. Time dependences of the bias applied between the PFM probe and the bottom electrode **(a)**, the vertical displacement $u_z$ of the surface below the probe apex **(b)** and the average polarization $\bar{P}_z^{aver}$ **(c)** in the heterostructure "probe – $Al_{1-x}Sc_xN/AlN/Al_{1-x}Sc_xN$ three-layer – bottom electrode". Bias dependences of the surface displacement $u_z$ **(d)**, the polarization $P_z$ averaged over a larger volume $V_c$ **(e)** and a smaller volume $V_t$ **(f)**. The distribution of polarization **(g)** in the cross-section of the three-layer at the moments of time numbered from "1" to "10" shown by the pointers in (a) – (f). The thickness of the $Al_{0.73}Sc_{0.27}N$ layers is 10 nm and the thickness of AlN layer is 20 nm, other parameters and conditions are the same as in **Fig. 2**.

### B. Proximity effect of polarization switching in the wurtzite bilayers and three-layers

It is very important to note that we observe the proximity effect of the hysteresis loops shown in **Figs. 2(f)-5(f)**, discovered earlier by Skidmore et al. [1] for electroded films, including AlN/$Al_{1-x}Sc_xN$ bilayers and three-layers. Indeed, the "otherwise irreversible" polarization of the AlN layer(s) switches simultaneously with the ferroelectric polarization of the $Al_{1-x}Sc_xN$ layer(s). As can be seen from the images "1" – "10" in **Fig. 2(g) 3(g), 4(g)** and **5(g)**), the proximity switching is observed in the AlN layers, where the ferroelectric nanodomains emerge and/or intergrow.

Note, the proximity effect of polarization switching occurs via the nanodomain nucleation for the probe-electrode geometry, because the external electric field is localized under the biased PFM probe in the considered case. To understand the features of the proximity switching in the studied AlN/$Al_{1-x}Sc_xN$ multilayers, one should consider that the local switching of polarization (as well as corresponding vertical displacement of the surface under the probe) occur in six stages, which are described below using an example of the AlN/$Al_{1-x}Sc_xN$ bilayer. Namely:

- Pre-nucleation stage I. The polarization switching does not occur at low voltages, only small local changes in polarization and surface displacement are observed (see the image "1" **Figs. 2(g)**). The images labeled "4", "7" and "10" in **Fig. 2(g)** correspond to the same stage but for increasing applied bias.
- Nucleation stage II. When the bias exceeds the coercive value, a nanodomain nucleation appears between the time moments "1" and "2", shown by inset labeled as "d.n." in **Figs. 2(g)**. Notably that the small nucleus rapidly transforms into the next stage. The time between the moments "4" and "5" correspond to the same nucleation stage as between "1" and "2", but realized for a larger applied bias. Note that the domain nucleus appears near the surface of the AlN layer under the probe, because the probe electric field is maximal in the spatial region (see **Fig. 1(a)**) and the direction of depolarization field favors the domain formation in this case (see **Fig. 1(f)**). It should be emphasized that the domain nucleation becomes possible in the unswitchable AlN layer because its "effective" double-well ferroelectric potential is renormalized to lower the steepness of the switching barrier due to the proximity of the switchable $Al_{1-x}Sc_xN$ layer [24]. The "agent" of the renormalization is the depolarization field, which is co-directed with the spontaneous polarization of $Al_{1-x}Sc_xN$ layer and counter-directed with the spontaneous polarization of the AlN layer (see **Fig. 1(f)**), as thus the field favors the domain formation in the AlN layer.
- Vertical growth stage III. The nanodomain elongates, rapidly grows vertically through both layers, reaches the bottom electrode, and forms a cylindrical domain with uncharged domain walls (see the images "2", "5" and "8" in **Fig. 2(g)**). The stage is rapid, namely it lasts the times comparable with a very small Landau-Khalatnikov (LK) relaxation time, due to the proximity effect. Note that the period of applied voltage is four orders of magnitude higher than the LK relaxation time in the considered case.

- Lateral growth stage IV. The cylindrical domain growths in the transversal direction with the bias increase (after the time moment "2", "5" and "8" in **Figs. 2(g)**). The transverse growth is much slowly (i.e., lasts the times comparable with the hundreds of LK times) in comparison with the rapid vertical growth.
- Back-switching stage V. When the probe bias decreases, the domain size does not decrease at once due to the pinning effect associated with the finiteness of the numerical grid. Only when the bias changes its sign and overcomes the critical value, a new domain nucleus appears inside the existing cylindrical domain (see the images "6" and "9" in **Figs. 2(g)**).
- Complete switching stage VI. When the magnitude of the bias increases further, the walls of the "nested" domain collapse and the complete switching occurs under the probe (see the images "7" and "10" in **Figs. 2(g)**).

When changing the sequence of the layers, e.g., when swapping the AlN and $Al_{1-x}Sc_xN$ layers, the width of hysteresis loop increases significantly (compare **Fig. 3** with **Fig. 2**). Looking at the free energy form shown in **Fig. 1(g)**, one might expect that the negative and positive coercive fields should be much higher for the AlN/$Al_{1-x}Sc_xN$ bilayer than the fields for the $Al_{1-x}Sc_xN$/AlN bilayer due to the following reasons. The nucleus of a new domain arises under the surface of the layer that is in direct contact with the charged probe (because the probe electric field is maximal under the surface), which is the $Al_{1-x}Sc_xN$ layer in the $Al_{1-x}Sc_xN$/AlN bilayer (see **Figs. 3(g)**) and the AlN layer in the AlN/$Al_{1-x}Sc_xN$ bilayer (see **Figs. 2(g)**). Since the potential wells of the AlN are much deeper than that of $Al_{1-x}Sc_xN$, the nucleation threshold bias is expected to be smaller for $Al_{1-x}Sc_xN$. However, our calculations showed the opposite trend: we reveal that a significantly larger external probe bias is needed for domains nucleation in the $Al_{1-x}Sc_xN$/AlN bilayer in comparison with the bias for the AlN/$Al_{1-x}Sc_xN$ bilayer (see **Table III** and compare **Figs. 2(e)-2(f)** with **Figs. 3(e)-3(f)**).

This opposite trend is explained due to the proximity effect [1] in the multilayers, being a direct consequence of the field inversion effect (such as shown in **Figs. 1(f)**). Namely, the internal electric field depolarizes the "strong" non-ferroelectric AlN layer with a large spontaneous polarization and polarizes the "weaker" ferroelectric $Al_{1-x}Sc_xN$ layer with a smaller spontaneous polarization [24]. When z-component of the PFM probe electric field is opposite to the direction of the spontaneous polarization in the layers, the internal depolarization field will be either subtracted (in the $Al_{1-x}Sc_xN$ layer) or added (in the AlN layer) to the external field. In conclusion, the coercive bias is smaller for the AlN/$Al_{1-x}Sc_xN$ bilayer than those in the $Al_{1-x}Sc_xN$/AlN bilayer. Note, that the effective double-well LGD potential is changed by a proximity effect in both layers [24], and that the physical reason of the LGD potential change is the counter-directed depolarization field induced by the difference of the layer spontaneous polarizations.

The conclusion maintains its validity for the three-layers, AlN/$Al_{1-x}Sc_xN$/AlN and $Al_{1-x}Sc_xN$/AlN/$Al_{1-x}Sc_xN$, whose local polarization reversal and effective piezoelectric response is shown in **Fig. 4** with **Fig. 5**, respectively. The larger coercive biases and the wider polarization loop correspond to the $Al_{1-x}Sc_xN$/AlN/$Al_{1-x}Sc_xN$ three-layer (see **Table III**). Moreover, the features of the local polarization reversal and surface displacement under the probe are very similar for the AlN/$Al_{1-x}Sc_xN$ bilayers and the AlN/$Al_{1-x}Sc_xN$/AlN three-layers; as well as for the $Al_{1-x}Sc_xN$/AlN bilayers and the $Al_{1-x}Sc_xN$/AlN/$Al_{1-x}Sc_xN$ three-layers. This observation is valid because the total thickness of the AlN and $Al_{1-x}Sc_xN$ layers is the same for the bilayer and three-layer structures (e.g., for the $h_1 = h_2$ =20 nm for the cases shown in **Figs 1-5**). However, the $Al_{1-x}Sc_xN$/AlN/$Al_{1-x}Sc_xN$ three-layer exhibits lower coercive biases compared to the $Al_{1-x}Sc_xN$/AlN bilayer (see **Table III**). This happens because the thickness of the top $Al_{1-x}Sc_xN$ layer in the three-layer is twice as small as that in the bilayer. Thus, the spike-like nanodomain, that nucleates under the probe apex, intergrows the twice thinner layer of $Al_{1-x}Sc_xN$, where the direction of the "net" depolarization field precludes its growth. Then it penetrates to the thicker central AlN layer, where the depolarization field helps its further growth. When the nanodomain reaches the bottom $Al_{1-x}Sc_xN$ layer, where the depolarization field counteracts its growth again, its sizes are enough large for the domain breakdown effect [58], and the lateral growth starts rapidly. Therefore, the coercive bias is mostly determined by the thickness of the top $Al_{1-x}Sc_xN$ and AlN layers. For the same reason the AlN/$Al_{1-x}Sc_xN$/AlN three-layers show higher coercive biases than the AlN/$Al_{1-x}Sc_xN$ bilayers (see **Table III**). Here the thickness of the top AlN layer in the three-layer is twice as small as that in the bilayer. Thus, the nanodomain, that nucleates under the probe apex, intergrows the twice thinner layer of AlN, where the direction of the "net" depolarization field co-acts its growth, then it penetrates to the thicker central $Al_{1-x}Sc_xN$ layer, where the depolarization field counteracts its further growth. Again, the coercive bias is mostly determined by the thickness of the top AlN and $Al_{1-x}Sc_xN$ layers. Moreover, it can be shown that the "net" depolarization fields calculated in the ground-state (at $U = 0$) are the same for the three-layers and bilayers (see **Supplement S1** in **Supplementary Materials** [57]).

### C. Size effect of the proximity switching in wurtzite multilayers

**Figure 6** shows the hysteresis loops of local polarization $\bar{P}_z^{local}$, average polarization $\bar{P}_z^{aver}$ and surface vertical displacement $u_z$ calculated when $h_2$ changes from 0 ($Al_{1-x}Sc_xN$ layer) to $h$ (AlN layer); at that the condition of constant sum $h_1 + h_2 = h$ (e.g., for $h$ =40 nm) is valid. It is seen from the figure that the coercive bias increases from 20 V to 30 V with increase in $h_2$ from 0 nm to 40 nm.

Interesting that the step-like features, which are almost absent (for $h_1 \leq 10$ nm) or relatively small (for 10 nm$< h_1 \leq 20$ nm) at the hysteresis loops of $\bar{P}_z^{local}$, are much larger and/or more pronounced at the hysteresis loops of $\bar{P}_z^{aver}$. The position and size of the step-like features of $\bar{P}_z^{aver}$ and $u_z$ correlate at positive biases. At the same time the position and size of the step-like features of $\bar{P}_z^{local}$ and $u_z$ correlate at negative biases. This observation proves the urgency to consider both polarizations, $\bar{P}_z^{local}$ and $\bar{P}_z^{aver}$, for the explanation of the $u_z$ behavior.

It is worth noting that the negative and positive coercive biases are indifferent to the way of averaging, being the same for $\bar{P}_z^{local}$, $\bar{P}_z^{aver}$ and $u_z$ loops (compare the first, the second and the third columns in **Fig. 6**). The coercive field of the 40-nm thick AlN layer (about 7.5 MV/cm) is higher than the field of electric breakdown (about 6 MV/cm). The average coercive field of the 40-nm thick $Al_{1-x}Sc_xN$ layer (about 5 MV/cm) is smaller than the breakdown field. Remarkably, both coercive fields calculated in the probe-electrode geometry are much smaller than the thermodynamic coercive fields in the capacitor geometry, which are 26 MV/cm for the AlN and 9.5 MV/cm for $Al_{1-x}Sc_xN$ [24]. The coercive fields of the $Al_{1-x}Sc_xN$/AlN multilayers varies in the range 5 – 7.5 MV/cm in dependence on the layer thickness ratio $h_1/h_2$ for the fixed probe size ($R = 5$ nm in the considered case). The strong reduction of the coercive fields occurs in the probe-electrode geometry because the biased probe acts as an external charged defect. The external charged defect appeared strong in comparison with a point charged defects in the bulk (considered in Ref. [24]).

It is interesting that the coercive bias of the AlN/$Al_{1-x}Sc_xN$ bilayers with $h_1 \geq 20$ nm is still about 20 V for the red hysteresis loops, which corresponds to the coercive field about 5 MV/cm (see **Figs. 6(a1)-(c3)**). The visible increase of the coercive bias is observed at $h_1 \leq 10$ nm (see **Figs. 6(d1)-(e3)**). The observation allows us to conclude that the coercive biases (as well as other features of the local polarization reversal) depend nonlinearly on the thickness ratio $h_1/h_2$. The nonlinear dependence of the loop parameters on layer thickness reflects the nonlinear origin of the proximity effect.

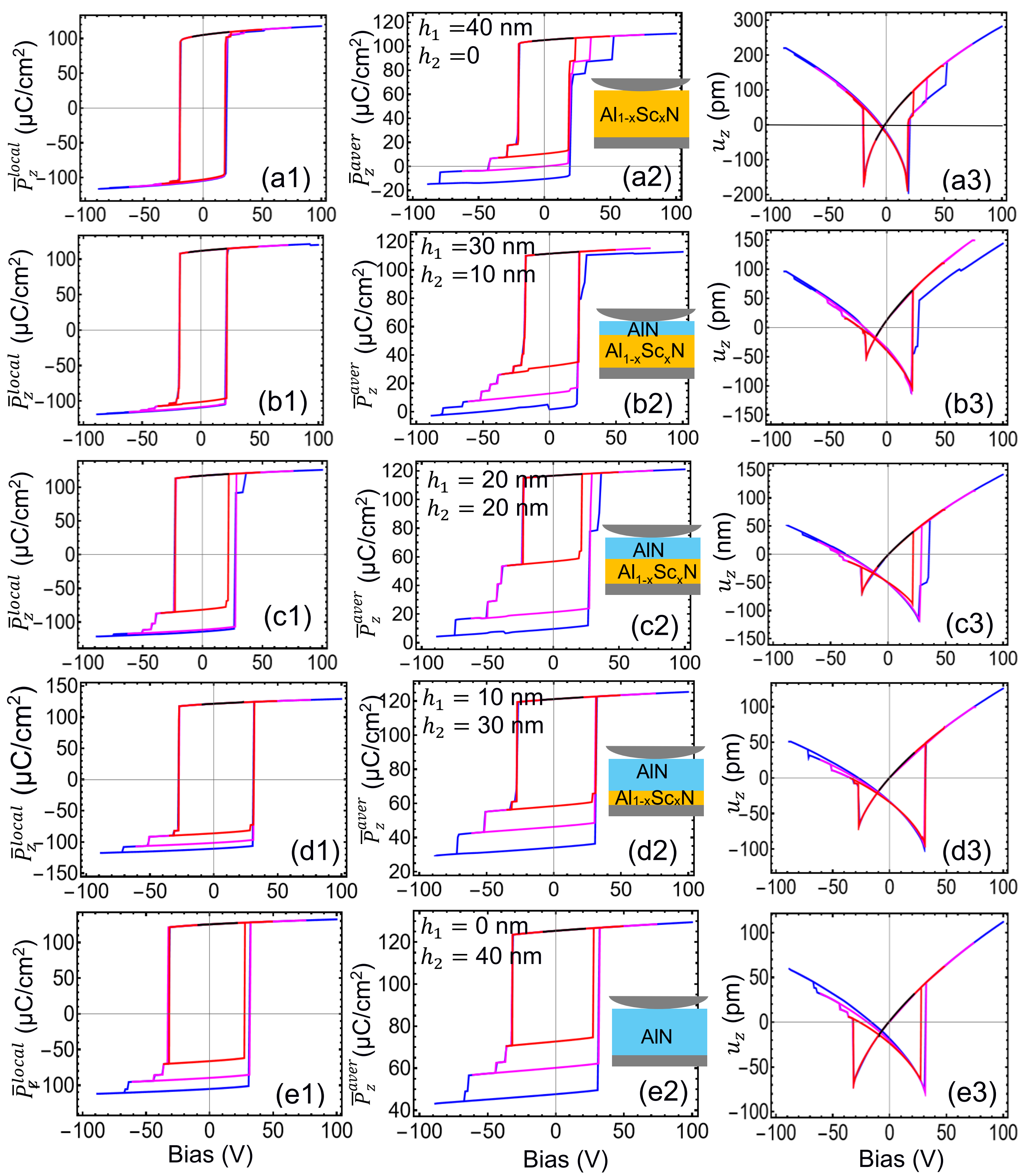

**FIGURE 6.** Bias dependences of the local polarization $\bar{P}_z^{local}$ **(a1)-(e1)**, averaged polarization $\bar{P}_z^{aver}$ **(a2)-(e2)**, and vertical surface displacement $u_z$ **(a3)-(e3)** in the heterostructure "probe – AlN/$Al_{0.73}Sc_{0.27}N$ bilayer – bottom electrode". Thicknesses $h_1$ of the $Al_{0.73}Sc_{0.27}N$ layer and the thickness $h_2$ of the AlN layer, are indicated on the plots (a2)-(e2). Other parameters and conditions are the same as in **Fig. 2**.

The dependences of the negative ($U_c^-$) and positive ($U_c^+$) coercive biases on the thickness $h_1$ can be analyzed from **Fig. 7(a)**. Red and blue curves in the plot correspond to the red and blue loops in **Fig. 6**, respectively. Note that the coercive biases are almost the same for the loops of $\bar{P}_z^{local}$, $\bar{P}_z^{aver}$ and $u_z$. The dependences are nonlinear and tend to saturate at $h_1 > 30$ nm. The asymmetry of coercive biases and insignificant oscillations of the red curves originate from the step-like features at the hysteresis loops related with the jump-like changes in the domain structure under the probe. The negative coercive bias $U_c^-$ in the same for red, magenta and blue loops, while the positive coercive bias $U_c^+$ is a bit larger for the blue loop.

The thickness dependences of the minimal ($P_r^-$) and maximal ($P_r^+$) values of remanent polarization are shown in **Fig. 7(b)**. Red and blue curves in the plot correspond to the red and blue loops of $\bar{P}_z^{local}$; and the green curve corresponds to the blue loops of $\bar{P}_z^{aver}$ in **Fig. 6**, respectively. The dependence of $(P_r^+)^{local}$ and $(P_r^-)^{local}$ vs. the thickness $h_1$ are relatively weak; meanwhile dependence of $(P_r^-)^{aver}$ vs. the thickness $h_1$ is relatively strong. The difference in the thickness dependences of $(P_r^\pm)^{local}$ and $(P_r^-)^{aver}$ can be explained as following: the magnitude of $(P_r^\pm)^{local}$ is mainly determined by the polarization distribution under the probe, and the magnitude of $(P_r^-)^{aver}$ is mainly determined by the polarization distribution far from the probe.

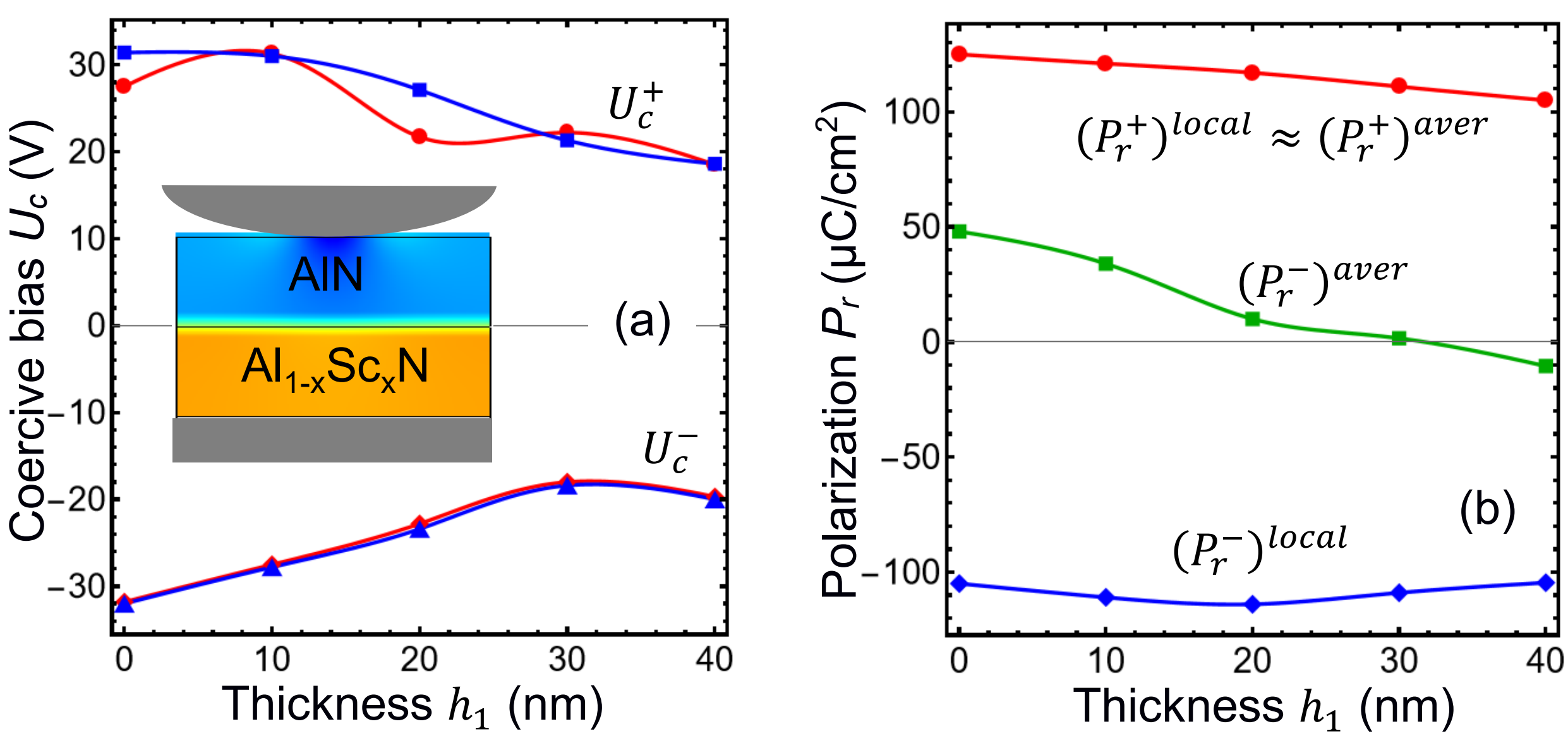

**FIGURE 7. (a)** The thickness dependences of the left ($U_c^-$) and right ($U_c^+$) coercive biases in the heterostructure "probe – AlN/$Al_{1-x}Sc_xN$ bilayer – bottom electrode". Red and blue curves correspond to the red and blue loops in **Fig. 6**, respectively. At that the coercive biases are the same for $\bar{P}_z^{local}$, $\bar{P}_z^{aver}$ and $u_z$. **(b)** The minimal ($P_r^-$) and maximal ($P_r^+$) values of remanent polarization in the same heterostructure. Red and blue curves correspond to the red and blue loops of $\bar{P}_z^{local}$, respectively, shown in the panels **(a1)-(e1)** in **Fig. 6**. The green curve corresponds to the blue loops of $\bar{P}_z^{aver}$, shown in the panels **(a2)-(e2)** in **Fig. 6**. Thicknesses $h_1$ of the $Al_{0.73}Sc_{0.27}N$ layer changes from 0 to 40 nm; and the thickness $h_2$ of the AlN layer is $40 - h_1$ (in nm). Other parameters and conditions are the same as in **Fig. 2**.

Above we considered bilayers and three-layers only. We expect that proximity ferroelectricity should exist in the multilayers with the total thickness of otherwise non-switchable layers comparable or smaller than the total hickness of switchable layers. In this case the size effect of proximity switching depends on the thickness ratio of the individual layers $h_1/h_2$, total thickness $h$ and on the tip-surface contact radius $R$. In particular, if the total thickness of the four-layer system is fixed (e.g., 40 nm as above), and one consider four alternating 10-nm thick layers, the behavior of local and average polarization will look very similar to those of three-layer system, but the horizontal assymetry of the loops and the dependence of coercive biases on the sequence of the layers under the probe will be smaller. This happens because the probe field in the second and third layers of the four-layer system is larger than in the case of three-layer system. Next, if we consider e.g., an ten-layer system with the total thickness $h$ =40 nm and 4-nm thick alternating AlN and AlScN layers, the dependence of coercive biases on the sequence of the layers under the probe becomes very small for $R$ =5 nm, because $h_i < R$ and the difference in the probe field is small in the two upper layers. The conditions $h_i \ll R$ and $h \gg R$ lead to the symmetric proximity-induced polarization switching in the multilayer, and we postpone the consideration of the case to further study.

## 5. Summary

Using the LGD thermodynamic approach, we explored the proximity effect of local piezoelectric response and polarization reversal in wurtzite ferroelectric multilayers. We analyze the probe-induced nucleation of nanodomains, the features of local polarization hysteresis loops and coercive fields in the $Al_{1-x}Sc_xN$/AlN bilayers and three-layers. The regimes of "proximity switching" (where the multilayers collectively switch) and the regime of "proximity suppression" (where they collectively do not switch) are the only possible regimes in the probe-electrode geometry. This result is reminiscent of the case of wurtzite multilayers sandwiched between flat electrodes.

We predict that the coercive bias, horizontal asymmetry and step-like features of the local polarization and piezoresponse hysteresis loops depend significantly on the sequence of the layers with respect to the probe. We find that the negative and positive coercive biases (and hence the loop width) are significantly smaller when the AlN layer is under the probe. This result is explained due to the proximity effect in the multilayers, being a direct consequence of the field inversion effect. The physical mechanism of the proximity ferroelectricity in the local probe geometry is a depolarizing electric field determined by the polarization of the layers and their relative thickness in a self-consistent manner. Namely, the internal electric field, whose direction is "inverted" in the layer(s) with the larger spontaneous polarization, depolarizes the non-ferroelectric AlN layer with a large spontaneous polarization and polarizes the ferroelectric $Al_{1-x}Sc_xN$ layer with a smaller spontaneous polarization. Thus, the depolarization field renormalizes the double-well ferroelectric potential to lower the steepness of the switching barrier in the nominally "unswitchable" polar layers and hence reduces the coercive field.

The coercive fields calculated by the FEM in this work in the probe-electrode geometry, are much smaller than the thermodynamic coercive fields calculated earlier in the capacitor geometry [24]. In accordance with our calculations, the coercive field of the $Al_{1-x}Sc_xN$ layer (about 5 MV/cm) is smaller than the breakdown field. The coercive field of the AlN layer (about 7.5 MV/cm) is higher than the field of electric breakdown (about 6 MV/cm). The coercive fields of the $Al_{1-x}Sc_xN$/AlN multilayers varies in the range 5 – 7.5 MV/cm, dependent on the layer thickness ratio $h_1/h_2$ for the fixed probe size ~ 5 nm. The strong reduction of the coercive fields occurs in the probe-electrode geometry because the biased probe acts as an external charged defect, whose strength, with an increase in bias increase, is comparable to or higher than that of the point charge defects located in the bulk of the layers. The ability of the tip-based proximity switching to differentially switch multilayers, based on the order of the layers, provides a powerful knob for selective domain engineering. The ability to locally pattern domain structures in previously unswitchable ferroelectrics such as AlN by exploiting the proximity effect in combination tip-based switching techniques opens up new avenues for domain-based nanoengineered ferroelectric devices for memory, actuation and nonlinear optics.

**Acknowledgements.** Authors are very grateful to the Reviewers for useful suggestions and stimulating discussions. The work is supported by the DOE Software Project on “Computational Mesoscale Science and Open Software for Quantum Materials”, under Award Number DE-SC0020145 as part of the Computational Materials Sciences Program of US Department of Energy, Office of Science, Basic Energy Sciences. The work of E.A.E. is also funded by the National Research Foundation of Ukraine (project “Silicon-compatible ferroelectric nanocomposites for electronics and sensors”, grant N 2023.03/0127). The work of A.N.M. is also funded by the National Research Foundation of Ukraine (project “Manyfold-degenerated metastable states of spontaneous polarization in nanoferroics: theory, experiment and perspectives for digital nanoelectronics”, grant N 2023.03/0132). S.V.K. effort was supported by the UT Knoxville start-up funding. Obtained results were visualized in Mathematica 14.0 [61].

**SUPPLEMENTARY MATERIALS to the**

# "Tip-Based Proximity Ferroelectric Switching and Piezoelectric Response in Wurtzite Multilayers"

## SUPPLEMENT S1. The depolarization fields for capacitor geometry

The depolarization fields for capacitor geometry are the same for all bilayers and three-layers considered in our work. Indeed, using that $h_1 = h_3 = \frac{h_2}{2}\,\varepsilon_b^{(1)} = \varepsilon_b^{(3)} \neq \varepsilon_b^{(2)}$ and $\bar{P}_z^{(1)} = \bar{P}_z^{(3)} \neq \bar{P}_z^{(2)}$ for the considered three-layers, and $h_1 = h_2$, $\varepsilon_b^{(1)} \neq \varepsilon_b^{(2)}$ and $\bar{P}_z^{(1)} \neq \bar{P}_z^{(2)}$for the considered bilayers, we obtained the depolarization field in the three-layers

$$E_z^{(1)}(x,y,z,t) = -\frac{P_z^{(1)}-\bar{D}}{\varepsilon_0\varepsilon_b^{(1)}} \equiv -\frac{P_z^{(1)}-\bar{P}_z^{(2)}}{\varepsilon_0\left(\varepsilon_b^{(1)}+\varepsilon_b^{(2)}\right)}, \tag{S1a}$$

$$E_z^{(2)}(x,y,z,t) = -\frac{P_z^{(2)}-\bar{D}}{\varepsilon_0\varepsilon_b^{(2)}} \equiv -\frac{P_z^{(2)}-\bar{P}_z^{(1)}}{\varepsilon_0\left(\varepsilon_b^{(1)}+\varepsilon_b^{(2)}\right)}, \tag{S1b}$$

$$E_z^{(3)}(x,y,z,t) = -\frac{P_z^{(3)}-\bar{D}}{\varepsilon_0\varepsilon_b^{(3)}} \equiv -\frac{P_z^{(1)}-\bar{P}_z^{(2)}}{\varepsilon_0\left(\varepsilon_b^{(1)}+\varepsilon_b^{(2)}\right)}. \tag{S1c}$$

The fields (S1) are the same as for the bilayers, because the average electric displacement

$$\bar{D} = \frac{\left(\frac{h_1}{\varepsilon_b^{(1)}}\bar{P}_z^{(1)}+\frac{h_2}{\varepsilon_b^{(2)}}\bar{P}_z^{(2)}+\frac{h_3}{\varepsilon_b^{(3)}}\bar{P}_z^{(3)}\right)}{\left(h_1/\varepsilon_b^{(1)}\right)+\left(h_2/\varepsilon_b^{(2)}\right)+\left(h_3/\varepsilon_b^{(3)}\right)} = \frac{\left(\frac{\bar{P}_z^{(2)}}{\varepsilon_b^{(2)}}+\frac{\bar{P}_z^{(1)}}{\varepsilon_b^{(1)}}\right)}{\left(1/\varepsilon_b^{(1)}\right)+\left(1/\varepsilon_b^{(2)}\right)} \equiv \frac{\varepsilon_b^{(1)}\bar{P}_z^{(2)}+\varepsilon_b^{(2)}\bar{P}_z^{(1)}}{\varepsilon_b^{(1)}+\varepsilon_b^{(2)}} \tag{S2}$$

is the same for the considered bilayers and three-layers. These calculations also explains why the coercive fields in capacitor geometry are the same for the considered bilayers and three-layers and independent of the sequience of the layers.

## SUPPLEMENT S2. Additional Figures

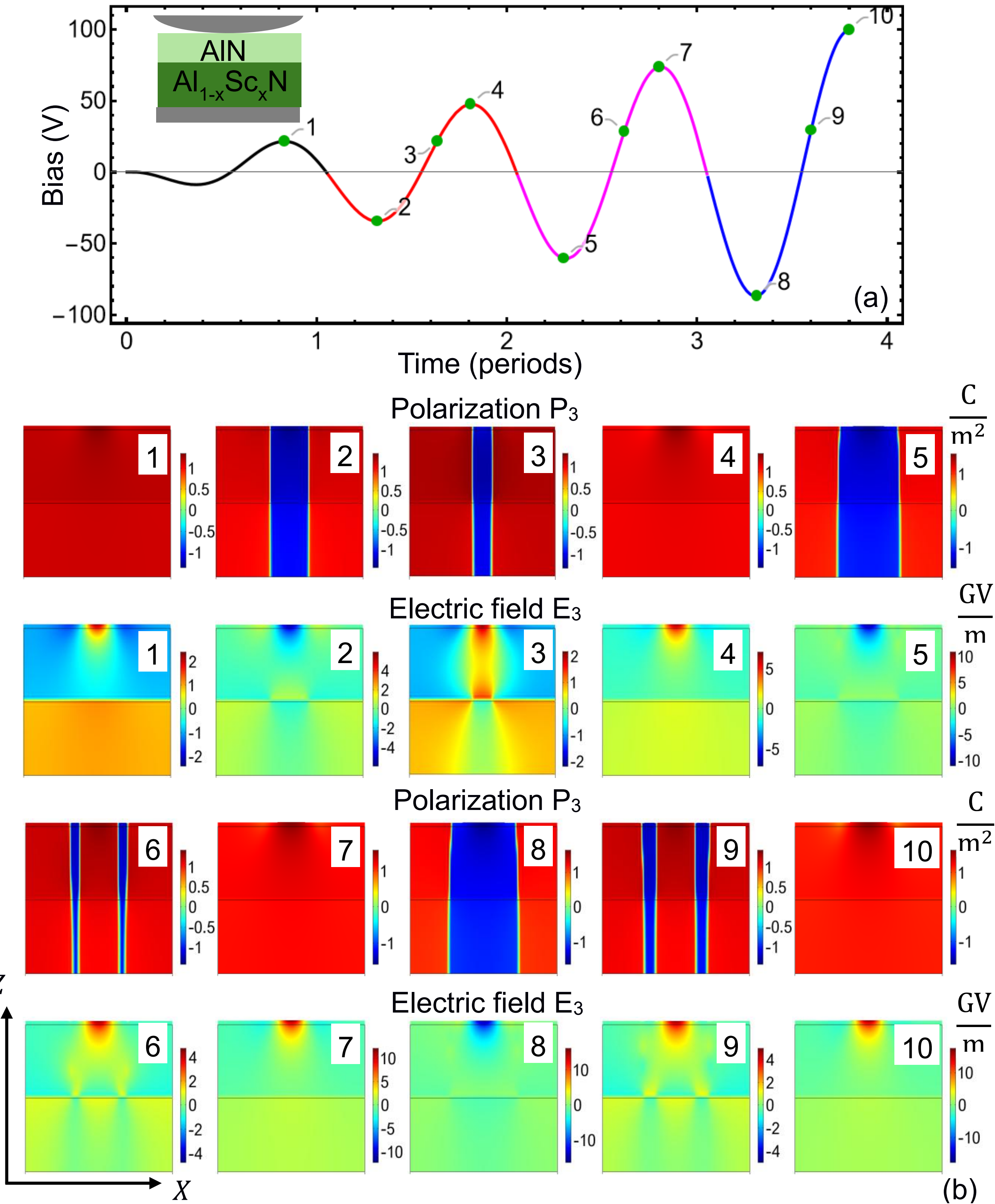

**FIGURE S1**. **(a)** Time dependence of the bias applied between the probe tip apex and the bottom electrode in the heterostructure "probe – bilayer AlN/$Al_{1-x}Sc_xN$– bottom electrode" for x=0.27. **(b)** The distribution of polarization component $P_z$ (the first and the third rows) and the electric field component $E_z$ (the second and the fourth rows) in the cross-section of the film at different moments of time with different values of the probe tip bias, numbered from 1 to 10 (see numbers with pointers in (a)). The thickness of AlN and $Al_{1-x}Sc_xN$ layers is 20 nm. LGD parameters and elastic constants are listed in **Tables I-II**. The initial distribution of polarization is a single-domain state with a randomly small fluctuation.

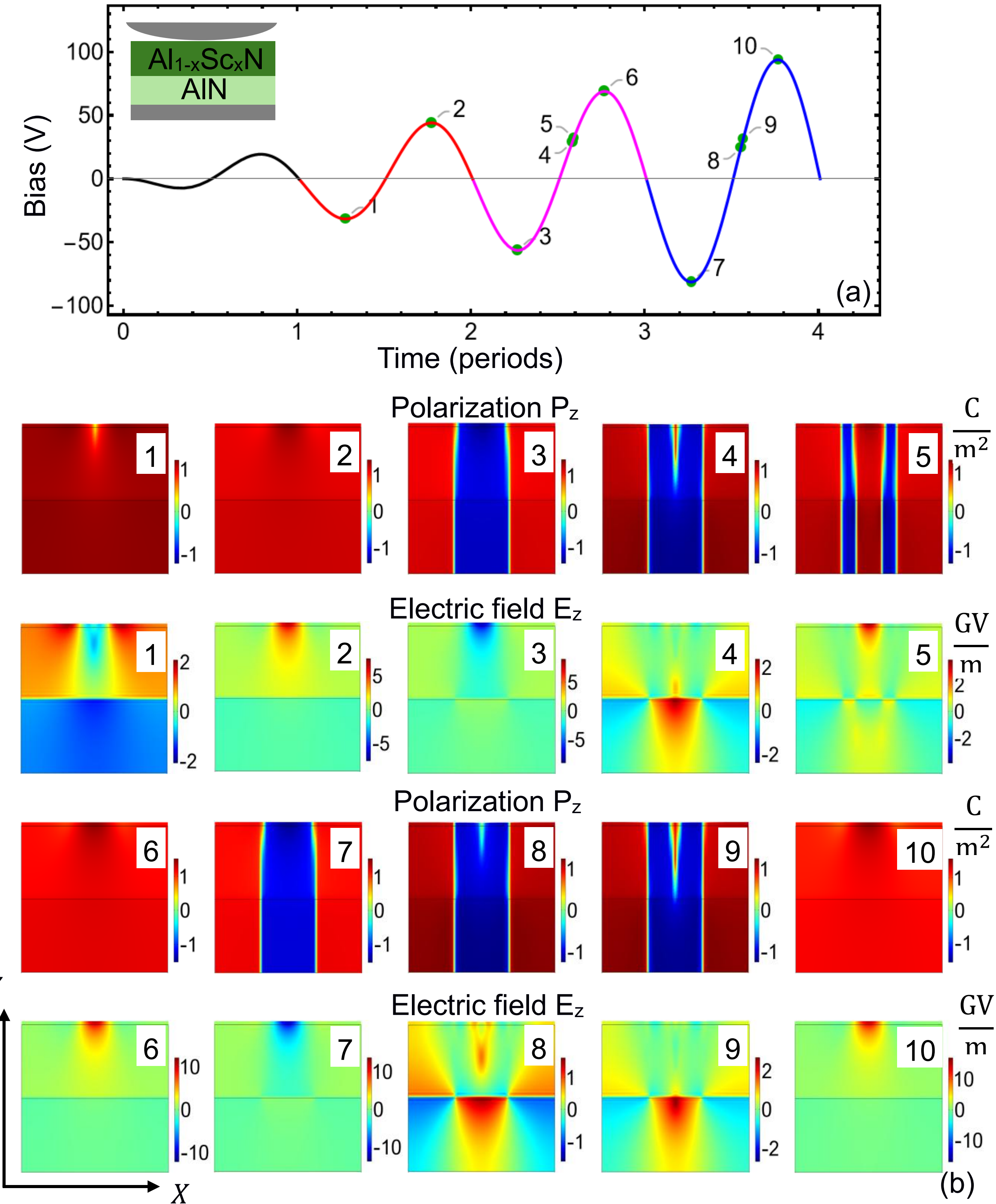

**FIGURE S2**. Time dependence of **(a)** the bias applied between the probe tip apex and the bottom electrode in the heterostructure “probe – bilayer $Al_{1-x}Sc_xN$/AlN– bottom electrode” for x=0.27. **(b)** The distribution of polarization component $P_z$ (the first and the third rows) and the electric field component $E_z$ (the second and the fourth rows) in the cross-section of the film at different moments of time with different values of the probe tip bias, numbered from 1 to 10 (see numbers with pointers in (a)). The thickness of AlN and $Al_{1-x}Sc_xN$ layers is 20 nm. LGD parameters and elastic constants are listed in **Tables I-II**. The initial distribution of polarization is a single-domain state with a randomly small fluctuation.

**Figures S3** and **S4** presenting the results of calculations starting from the initial polarization, consisting of the planar multidomain distribution with random functions, are shown below.

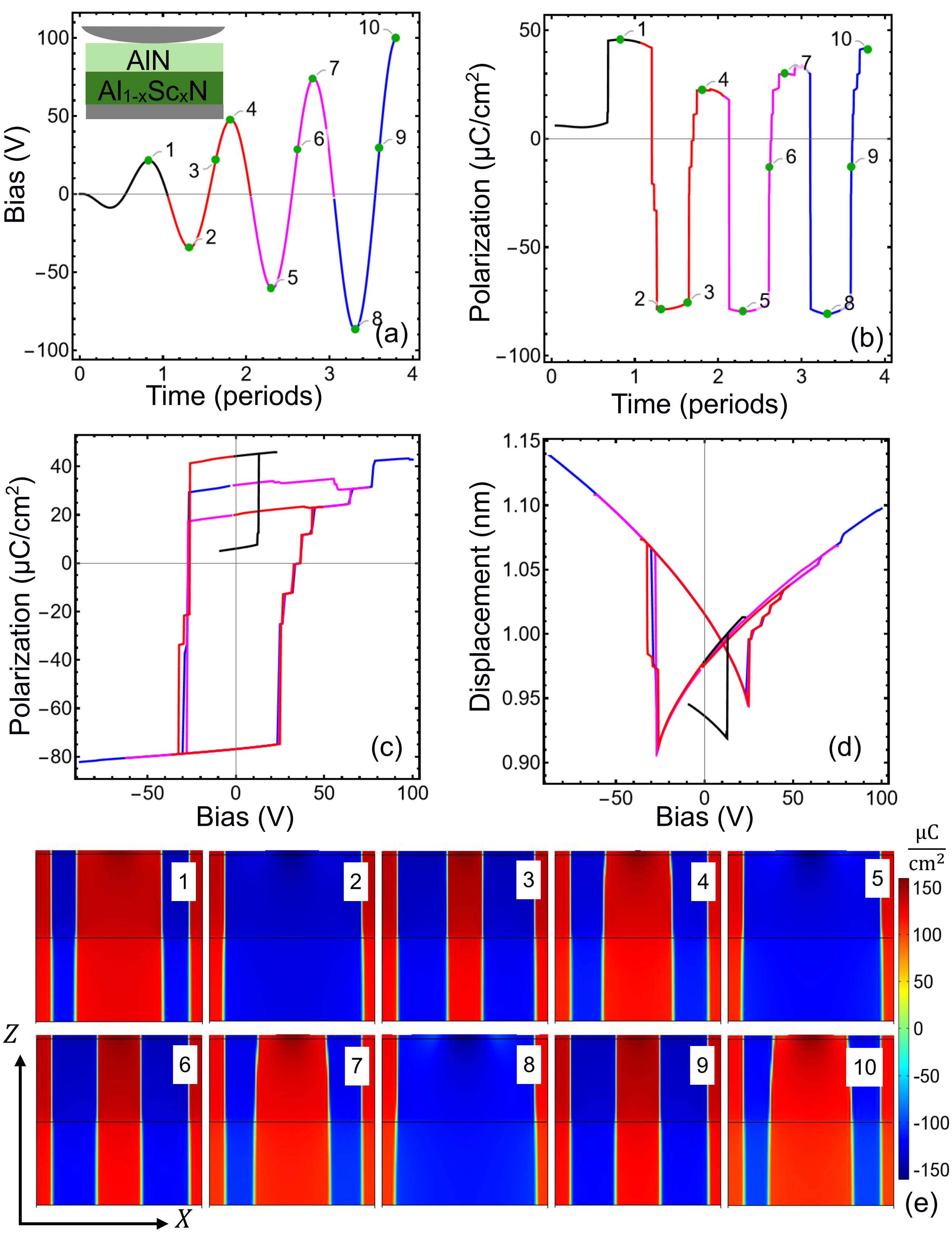

**FIGURE S3**. Time dependences of the bias applied between the PFM probe and the bottom electrode **(a)** and the average polarization $\bar{P}_z^{aver}$ of the bilayer $AlN/Al_{1-x}Sc_xN$ **(b)**. Bias dependences of the average polarization $\bar{P}_z^{aver}$ **(c)** and vertical

displacement $u_z$ of the surface below the probe apex **(d)** in the heterostructure "probe – bilayer AlN/$Al_{1-x}Sc_xN$ – bottom electrode". The distribution of polarization component $P_z$ **(e)** in the cross-section of the bilayer at the moments of time numbered from "1" to "10" shown by the pointers in (a) and (b). The thickness of the AlN and $Al_{1-x}Sc_xN$ layers is 20 nm, x = 0.27, the tip-surface contact radius $R = 5$ nm. LGD parameters and elastic constants are listed in **Tables I-II**. The initial distribution of polarization is a multi-domain state with randomly small fluctuations.

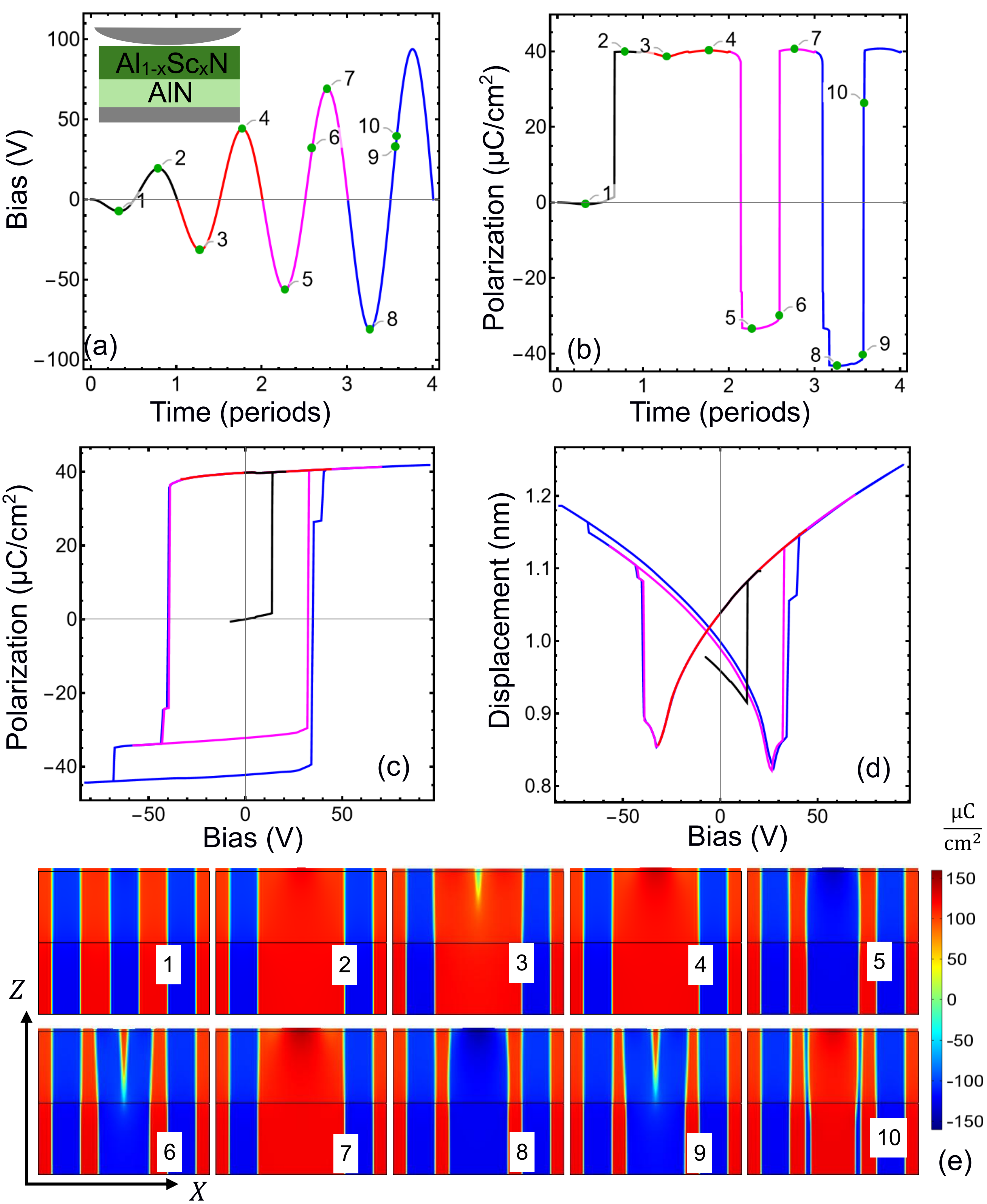

**FIGURE S4**. Time dependences of the bias applied between the PFM probe and the bottom electrode **(a)** and the average polarization $\bar{P}_z^{aver}$ of the bilayer $Al_{1-x}Sc_xN$/AlN **(b)**. Bias dependences of the average polarization $\bar{P}_z^{aver}$ **(c)** and vertical displacement $u_z$ of the surface below the probe apex **(d)** in the heterostructure "probe – bilayer $Al_{1-x}Sc_xN$/AlN – bottom

electrode". The distribution of polarization component $P_z$ **(e)** in the cross-section of the bilayer at the moments of time numbered from "1" to "10" shown by the pointers in (a) and (b). The thickness of the AlN and $Al_{1-x}Sc_xN$ layers is 20 nm, x = 0.27, the tip-surface contact radius $R = 5$ nm. LGD parameters and elastic constants are listed in **Tables I-II**. The initial distribution of polarization is a multi-domain state with randomly small fluctuations.

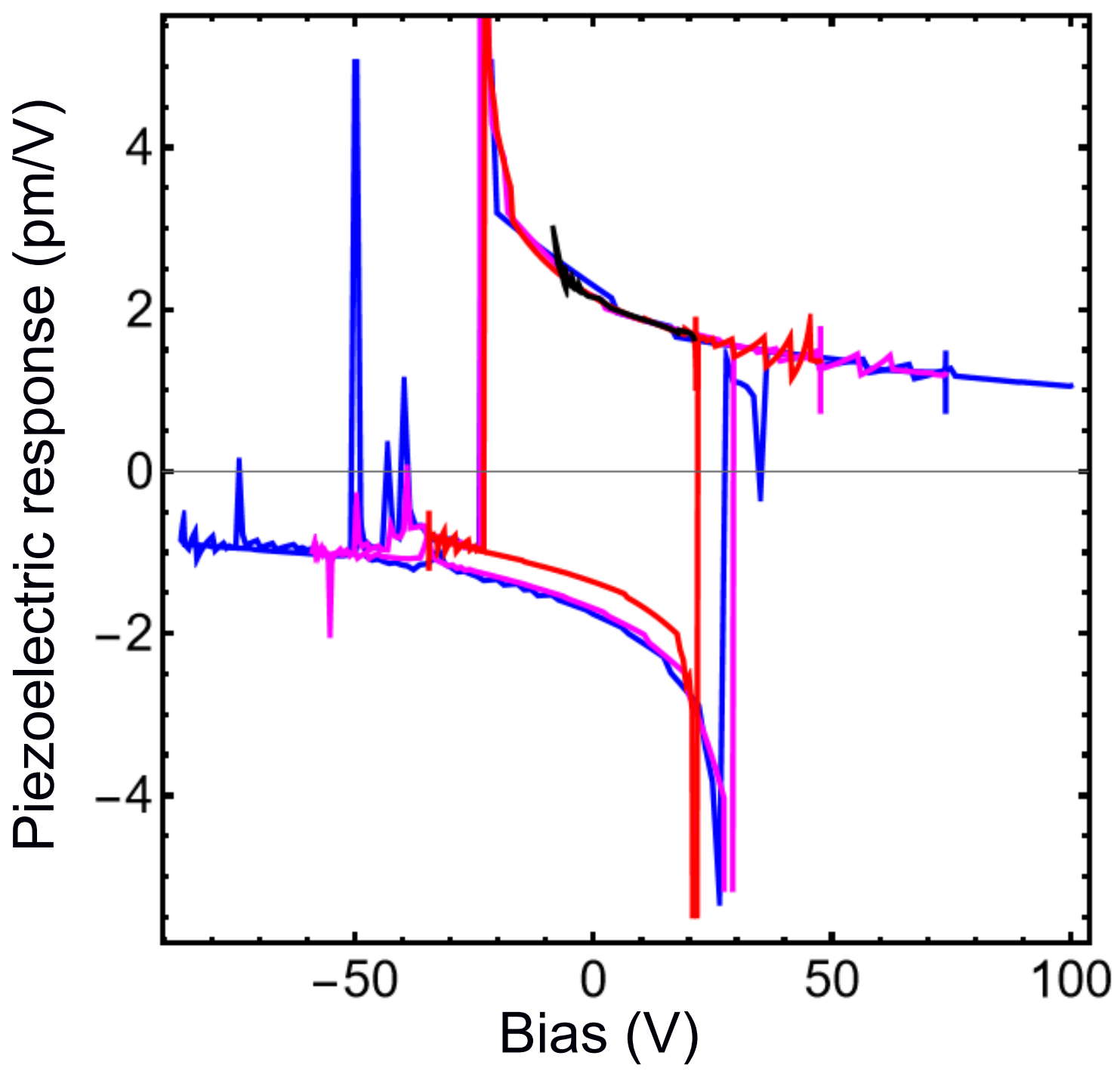

**FIGURE S5**. Effective piezoelectric response of the heterostructure "probe – AlN/$Al_{0.73}Sc_{0.27}N$ bilayer – bottom electrode". Other parameters and conditions are the same as in **Fig. 2.** Unfortunately, the FEM results are too noisy, and we failed to reduce the noise. So, this figure is for SI only, meanwhile the surface displacements are shown in the main text.